\documentclass[prd,twocolumn,a4paper,superscriptaddress]{revtex4}
\usepackage{graphicx}
\usepackage{amsmath,amsfonts,amstext,amssymb,amsbsy,amsopn,amsthm,eucal}
\usepackage{dsfont}

\begin{document}

\newcommand{\be}{\begin{equation}}
\newcommand{\ee}{\end{equation}}
\newcommand{\bq}{\begin{eqnarray}}
\newcommand{\eq}{\end{eqnarray}}
\newcommand{\bsq}{\begin{subequations}}
\newcommand{\esq}{\end{subequations}}
\newcommand{\bc}{\begin{center}}
\newcommand{\ec}{\end{center}}
\newcommand\lsim{\mathrel{\rlap{\lower4pt\hbox{\hskip1pt$\sim$}}
    \raise1pt\hbox{$<$}}}
\newcommand\gsim{\mathrel{\rlap{\lower4pt\hbox{\hskip1pt$\sim$}}
    \raise1pt\hbox{$>$}}}
\newcommand{\nn}{\nonumber}
\newcommand{\bo}{\raise-1mm\hbox{\Large$\Box$}} 
\newcommand{\simpropto}{
\begin{array}{c}
\propto \\[-1.6ex] \sim
\end{array}}

\title{Mass Inflation in Brans-Dicke gravity}
\author{P.P. Avelino}
\email[Electronic address: ]{ppavelin@fc.up.pt}
\affiliation{Departamento de F\'{\i}sica da Faculdade de Ci\^encias
da Universidade do Porto and Centro de F\'{\i}sica do Porto, \\ 
Rua do Campo Alegre 687, 4169-007 Porto, Portugal}
\author{A.J.S. Hamilton}
\email[Electronic address: ]{Andrew.Hamilton@colorado.edu}
\affiliation{JILA and Dept. Astrophysical $\&$ Planetary Sciences, \\
Box 440, U. Colorado, Boulder, CO 80309, USA}
\author{C.A.R. Herdeiro$^1$}
\email[Electronic address: ]{crherdei@fc.up.pt}
\date{\today}
\begin{abstract}
A detailed non-linear analysis of the internal structure of
spherical, charged black holes that are accreting scalar matter is performed in the framework of the Brans-Dicke theory of gravity. We choose the lowest value of the Brans-Dicke parameter that is compatible with observational constraints. First, the homogeneous approximation is used. It indicates that mass inflation occurs and that the variations of the Brans-Dicke scalar inside the black hole, which could in principle be large in the absence of mass inflation, become small when mass inflation does occur. Then, a full non-linear numerical study of the black hole interior perturbed by a self-gravitating massless uncharged scalar-field is performed. We use an algorithm with adaptive mesh refinement capabilities. In this way, the changes in the internal structure of the black hole caused by mass inflation are determined, as well as the induced variations of the Brans-Dicke scalar, confirming, qualitatively, the indications given by the homogeneous approximation.
\end{abstract}

\maketitle

\section{\label{intr}Introduction}

It has long been known that the internal structure of an eternal Reissner-Nordstr\"om or Kerr black hole must be drastically altered by accretion. Firstly, a Reissner-Nordstr\"om or Kerr black hole has an inner horizon covering a gravitationally repulsive singularity, which is qualitatively different from a Schwarzschild black hole; it is unreasonable that the accretion of an infinitesimal amount of charged or spinning matter could lead to such a dramatic change in the black hole interior. Secondly, there are arguments, of both physical \cite{Penrose:1968ar} and mathematical nature \cite{Chandra}, indicating that the Cauchy horizon (i.e.\ inner horizon) of the eternal charged or rotating hole is unstable against small (linear) perturbations. The natural question is then, what is the endpoint of the instability? Answering it should shed some light on the internal structure of realistic astrophysical Kerr black holes. 

Studying general perturbations of a stationary axi-symmetric spacetime is a hard problem in general relativity \cite{Teukolsky:1973ha}. Thus, the aforementioned questions can be initially considered in the context of spherical perturbations of the Reissner-Nordstr\"om black hole, as a simple model for the more challenging realistic case of the Kerr metric.  This was the route followed by Poisson and Israel, in their seminal work where the phenomenon of mass inflation was unveiled \cite{Poisson:1989zz,Poisson:1990eh}: if ingoing and outgoing streams are simultaneously present near
the inner horizon, then relativistic counter-streaming between those streams leads to exponential growth of gauge-invariant measures such as the interior (Misner-Sharpe \cite{Misner}) mass,
the center-of-mass energy density, or curvature scalar invariants. Since this effect is causally disconnected from any external observers, the mass of the black hole measured by an outside observer
remains unchanged by the mass inflation going on in the interior. But this inflation phenomenon causes the spacetime curvature to grow to Planckian values on a spacelike hypersurface in the neighbourhood of the Cauchy horizon.

A lot of effort has been put in understanding detailed aspects of the mass inflation phenomenon, within general relativity. In their original work, Poisson and Israel gave analytical arguments to support that the exponential growth of the internal effective mass was sourced by the existence of both ingoing and outgoing fluxes -- a relativistic counter-streaming. According to these authors, the ingoing flux is infinitely blueshifted as it approaches the Cauchy horizon. But by itself this does not suffice to trigger mass inflation; an outgoing flux that induces a separation between the Cauchy and the inner apparent horizon is also necessary.

Various other numerical and analytical studies in the framework of 
general relativity have shown that mass 
inflation develops naturally when a massless scalar field is accreted by a 
spherical charged black hole \cite{Burko:1997zy,Burko:1997tb,Burko:1998jz,Burko:2002qr,Burko:2002fv,Oren:2003gp,Hansen:2005am}.
Perturbations of the scalar field propagate at the speed of light, and 
outgoing fluxes are produced as a result of the scattering of the ingoing
scalar field by the space-time curvature, leading again to counter streaming and to mass inflation. Yet another example has been provided in
\cite{Hamilton:2004av,Hamilton:2004aw} by studying the self-similar accretion 
of charged baryons into spherical
black holes, showing that mass inflation occurs if an additional neutral
matter component is present. In this case it is the relativistic
counter-streaming between ingoing neutral matter and outgoing charged
baryons that appears to be responsible for mass inflation. The physics of the relativistic counter-streaming that drives mass inflation inside black holes is further analysed in \cite{Hamilton:2008zz}.

But it has not been explored, so far, if  these dramatic events taking place inside a black hole still occur in a more general theory of gravity.  This is the programme we wish to start with the present paper.

To be concrete, we shall consider one of the simplest extensions of general relativity,  the Brans-Dicke (BD) theory 
of gravity \cite{Brans:1961sx}, a well known example of a 
scalar-tensor theory in which the gravitational interaction is mediated by a 
scalar 
field, in addition to the metric. BD theory is parametrized by one 
additional parameter, $\omega$, as compared to general relativity, which is 
obtained in the limit $\omega \to \infty$. Hence, this theory 
passes all available experimental tests \cite{Will:1993ns} if $\omega$ 
is large enough. The most stringent limits were obtained 
from precision measurements of the timing of signals from the Cassini 
mission which constrained $\omega$ to be larger than $40000$ 
\cite{Bertotti:2003rm}.

The key observation for the present study is quite simple. If BD theory is minimally coupled to the Maxwell field, such a theory admits as a solution the standard Reissner-Nordstr\"om black hole of Einstein-Maxwell theory, irrespective of the value of $\omega$. Indeed, the variations of the BD scalar field, $\Phi$, are sourced by the trace of the matter energy-momentum tensor $T$, 
\be
\Box \Phi = \frac{8 \pi T}{3+2 \omega} \ , \label{eqPhibd}
\ee
which is zero for a classical electromagnetic field. Thus, the BD scalar $\Phi$ can be taken as constant, and the theory reduces to Einstein-Maxwell with an extra (irrelevant) scalar field.

Matter perturbations with non-vanishing energy-momentum trace will, however, induce variations of the BD scalar. Since this scalar determines the gravitational coupling to matter, it influences a back reaction of matter on the geometry. A priori, there are reasons to expect these variations to be large. 
The first reason is that equation \eqref{eqPhibd} with negligible or no sources, specialised to a stationary spherically symmetric spacetime in spherical coordinates,
\be ds^2=g_{tt}(r)dt^2+g_{rr}(r)dr^2+r^2d\Omega_2 \ , \label{sphericalmetric} \ee 
yields, in a region where the metric coefficients have fast variations (this follows from \eqref{Phiwave} below)
\be \frac{d\Phi}{dr}\simpropto \sqrt{\frac{|g_{rr}|}{|g_{tt}|}} \ ; \label{propintro}\ee
thus, near a horizon, in particular near the Cauchy horizon, one could expect large variations of $\Phi$ if an (even tiny) source term or perturbation in $\Phi$ exists.
The second reason is that mass inflation could potentially lead
to a large right hand side of (\ref{eqPhibd}). Thus, two natural (and related) questions follow: (a) does mass inflation still occur? (b) are there significant variations of the BD scalar, and hence of the gravitational coupling to matter?

In this paper we shall answer the above two questions, by studying the space-time geometry and the dynamics of the 
matter fields in the interior of an accreting spherical, charged 
black hole in the framework of BD gravity. We shall see that, choosing the lowest value of $\omega$ compatible with observational constraints, mass inflation occurs much in the same way as in general relativity, and no substantial variations of the Brans-Dicke scalar are found. That $\Phi$ varies little is actually a surprising conclusion, in view of the above arguments. As we shall argue, it is actually the mass inflation phenomenon that, in our model, prevents larger $\Phi$ variations from being produced.

This paper is organised as follows. We start by introducing our model in section \ref{model}. In this section we discuss the equations of motion, assuming spherical symmetry, specify the initial conditions for a full numerical integration and discuss the numerical algorithm used. In section \ref{homog} we describe the homogeneous approximation, from which we extract both numerical and analytical information about the occurence of mass inflation and the variations of the Brans-Dicke scalar. In section \ref{resul} we perform a full non-linear numerical evolution of the black hole 
interior perturbed by a self-gravitating massless scalar-field, using both a compact and a non-compact perturbation. In section \ref{conc} we draw conclusions.

\section{\label{model} Model}

\subsection{\label{fequations} Field equations with spherical symmetry}

Consider the Brans-Dicke (BD) action:
\be
S=\int d^4 x {\sqrt {-g}} \left(\Phi R - \omega \frac{\Phi_{,\mu} \Phi^{,\mu}}{\Phi}  + 16 \pi {\cal L}_M \right) \ , 
\label{action}
\ee
where $R$ is the Ricci scalar, $\Phi$ is a real scalar field and $\omega$ is the 
Brans-Dicke parameter for which, throughout this paper, we shall take the minimal value consistent with observational constraints \cite{Bertotti:2003rm}, namely
\be \omega=40000 \ . \ee
${\cal L}_M$ is the matter Lagrangian, which we will assume 
to be the sum of the usual Maxwell contribution, ${\cal L}_F$, and the 
contribution from a self-gravitating real massless scalar field $\varphi$: 
\be
{\cal L}_M = {\cal L}_F + {\cal L}_\varphi = - \frac{F^2}{16\pi} - \frac{1}{8 \pi} 
\varphi_{,\alpha} \varphi^{,\alpha}\, \ , \label{matterlag}
\ee
where $F^2=F_{\alpha \beta} F^{\alpha \beta}$ and $F_{\alpha \beta}$ is the Maxwell tensor. Our signature choice 
is $-+++$. 
The energy-momentum tensor of the matter fields is given by
 ${}^M T_{\mu \nu} = {}^F T_{\mu \nu} + {}^\varphi T_{\mu \nu}$, where
\bq
{}^F T_{\mu \nu} &=& \frac{1}{4\pi}\left(F_{\mu \alpha} {F_\nu}^\alpha - \frac{1}{4} g_{\mu \nu} F^2\right)\ , \\
{}^\varphi T_{\mu \nu} &=& \frac{1}{4 \pi} \left(\varphi_{,\mu} \varphi_{,\nu} - \frac{1}{2} g_{\mu \nu} \varphi_{,\alpha} \varphi^{,\alpha}\right)\, \ .
\eq
The energy-momentum of the BD field is
\bq
8 \pi {}^\Phi T_{\mu \nu} &=& \frac{\omega}{\Phi}
(\Phi_{,\mu} \Phi_{,\nu}-\frac{1}{2} g_{\mu \nu} \Phi_{,\alpha} 
\Phi^{,\alpha})
\nonumber
\\
&+& \Phi_{;\mu \nu} - g_{\mu \nu} \Box \Phi \, \ .
\eq
The equations of motion derived from (\ref{action}) with (\ref{matterlag}) are the gravitational equations
\bq
G_{\mu \nu} &=& \frac{8 \pi}{\Phi} ({}^M T_{\mu \nu} + {}^\Phi T_{\mu \nu})\, \ , \label{einstein}
\eq
equation \eqref{eqPhibd} for the BD field,
with $T \equiv {}^M {T_\alpha}^\alpha$,
and the matter equations
\be
\label{eqphif}
\Box \varphi=0\, \ , \qquad d\star F=0 \ , 
\ee
where $\star$ denotes Hodge dual.

To solve the coupled system  (\ref{eqPhibd}), (\ref{einstein}) and (\ref{eqphif}), we take a spherically symmetric ansatz written in  double-null coordinates. Thus
\be
ds^2=-2e^{2\sigma(u,v)}du dv + r^2(u,v) d\Omega^2\, \ , \label{metric}
\ee
\be
\Phi=\Phi(u,v) \ , \ F=F_{uv}(u,v)du\wedge dv \ , \ \varphi=\varphi(u,v) \  , \ee
where $u$ and $v$ are taken to be ingoing and outgoing respectively. Under a gauge transformation of the form 
$u \rightarrow \mbox{function of } u$ and $v \rightarrow \mbox{function of } v$,
the coordinates $u$ and $v$ still preserve their null character; this will be of use later.

The Maxwell equations~(\ref{eqphif}) are simply solved to yield
\be
\label{maxwell1}
F_{u v}r^2e^{-2\sigma}= {\rm constant}= q \, \ ,
\ee
where
\be q=\frac{1}{8\pi}\oint F \ , \ee
is the electric charge. The electric field is therefore purely radial, as expected from spherical symmetry. The scalar field equation~(\ref{eqphif}) gives
\be
\varphi_{,uv} = -\frac{1}{r}\left(r_{,v} \varphi_{,u} - 
r_{,u} \varphi_{,v}\right) \ .
\label{scalar}
\ee

The $uu$, $vv$, $uv$ and $\theta\theta$ components of the Einstein equations~(\ref{einstein}), which are the only distinct non-trivial ones, give rise to the following equations
(the left hand sides these equations are components of the Einstein tensor
$G_{\mu\nu}$):
\be  \frac{4 r_{,u} \sigma_{,u} - 2 r_{,uu}}{r}=\frac{2(\varphi_{,u})^2+\Phi_{,uu}-2 \Phi_{,u} 
\sigma_{,u}}{\Phi} +w \frac{\Phi_{,u}^2}{\Phi^2}  \ , \label{einuu}\ee
\be  \frac{4 r_{,v} \sigma_{,v} - 2 r_{,vv}}{r}=\frac{2(\varphi_{,v})^2+\Phi_{,vv}-2 \Phi_{,v} 
\sigma_{,v}}{\Phi} +w \frac{\Phi_{,v}^2}{\Phi^2}  \ , \label{einvv}\ee
\begin{eqnarray} 
\frac{e^{2 \sigma} + 2 r_{,v} r_{,u} +2 r r_{,uv}}{r^2}=
 \frac{q^2}{r^4}\frac{e^{2 \sigma}}{\Phi}-\frac{\Phi_{,uv}}{\Phi}  \ \ \ \ \ \ \ \ \ \nn \\   \ \ \  \ \ \ \ \ \  -\frac{2 (r_{,u}\Phi_{,v}+r_{,v}\Phi_{,u})}{r\Phi} \ , \ \ \  \ \ \label{einuv} \end{eqnarray}
\begin{eqnarray}
-2  r (r_{,uv} + r \sigma_{,uv})=\frac{q^2}{ r^2}\frac{e^{2 \sigma}}{\Phi}+\frac{2}{\Phi} r^2  \varphi_{,u} \varphi_{,v} \ \ \ \ \ \ \nn \\+\frac{r }{\Phi}\left(r_{,u} \Phi_{,v}+r_{,v} \Phi_{,u}\right)
+ 
\frac{r^2 }{ \Phi^2}\left( \omega \Phi_{,u} \Phi_{,v}+2 \Phi \Phi_{,uv}\right)\ . \ \ \ \ \label{eintt}  \end{eqnarray}
Equations~\eqref{einuu}-\eqref{einvv} are two constraint equations,
whereas
equations~\eqref{einuv}-\eqref{bdscalar} are evolution equations.
In the former equations, the contribution from the energy-momentum of the scalar field is
\be T^{\varphi}_{uu}=\frac{(\varphi_{,u})^2}{4\pi} \ , \qquad T^{\varphi}_{vv}=\frac{(\varphi_{,v})^2}{4\pi} \ , \ee
respectively, which represent, physically, the flux of the scalar field through surfaces of constant $v$ and $u$, i.e.\ outflux and influx. We shall impose that, initially, only influx exists, equation~\eqref{ioutflux}. However, outflux is inevitably produced by scattering off the spacetime geometry.

Finally, the wave equation~(\ref{eqPhibd}) for the Brans-Dicke scalar field $\Phi$ is
\bq
\Phi_{,uv} &=& - \frac{1}{r}\left(r_{,v} \Phi_{,u} + 
r_{,u} \Phi_{,v}\right)  -\frac{2 \varphi_{,u} \varphi_{,v}}{3+2 \omega} \ .
\label{bdscalar}
\eq

\subsection{\label{Iconditions} Initial conditions}

Without loss of generality,
we take the initial unperturbed RN black hole to have mass
\be m_0=1 \ , \ee 
which is simply a choice of units.
Observational evidence
\cite{McClintock:2006xd,Miller:2009cw,Brenneman:2006hw,Miller:2008vc}
suggests that astronomical black holes have large spins in several cases.
We therefore take the unperturbed RN black hole to have a charge
near the extremal value of unity,
\be q=0.95 \ . \ee 
The corresponding radii $r_-$ and $r_+$ of the unperturbed inner and outer horizons are
\be r_-\simeq 0.69 \ , \qquad r_+\simeq 1.31 \ . \label{horizons} \ee
Note that the mass of the black hole will change with accretion, but not its charge.

\begin{figure}[t!]
\begin{picture}(0,0)(0,0)
\put(-40,120){$v=v_0=5$}
\put(30,110){$r=r_0+\frac{u}{4}\left(\frac{2}{r_0}\left(m_0-\frac{q^2}{2r_0}\right)-1\right)$}
\put(44,96){$\sigma=-\ln2/2$}
\put(54,84){$\Phi=1$}
\put(64,74){$\varphi=0$}
\put(135,60){$u=u_0=0$}
\put(130,40){$r=v$}
\put(120,30){$\sigma_{,v}=v(\varphi_{,v})^2/2$}
\put(110,20){$\Phi=1$}
\put(100,10){$\varphi_{,v}=A\sin^2\left(\pi\frac{v-v_0}{\Delta v}\right)$}
\put(60,20){$u$}
\put(83,20){$v$}
\end{picture}
\includegraphics[width=1.8in,keepaspectratio]{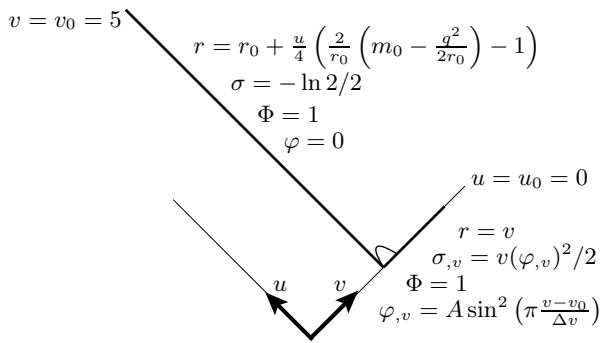}
\caption{\label{initialcond}  Initial conditions in the numerical integration box, defined by $u_0=0\le u<30$, $v_0=5\le v <20$,
for the case where the black hole is fed by a compact influx of scalar field $\varphi$.
The unperturbed black hole has mass and charge $m_0=1$, $q=0.95$.
The influx along the null segment $u = u_0$ starts at $v_0=5$, has amplitude $A=0.05$ or $A=0.1$, and endures for $\Delta v=1$.
The influx, depicted by a curve, is non-zero only for $v_0<v<v_0+\Delta v$. The scale in the $u$ and $v$ axes is different to make closer contact with the Carter-Penrose diagrams shown in Figs.~\protect\ref{figurecp} and \protect\ref{figurecpaccretion}.}
\end{figure}

The numerical integration is performed in double-null coordinates
$u$, $v$
over an integration box
$u_0 = 0 \leq u < 30$,
$v_0 = 5 \leq v < 20$.
Initial conditions are prescribed along the two null segments
$u = u_0$ and $v = v_0$.
Prior to $v = v_0$, the geometry is Reissner-Nordstr\"om.
We consider two cases,
one in which the black hole is fed with a compact influx of scalar field $\varphi$
that extends over a finite time,
and another in which the black hole is fed with a non-compact influx
that starts at a certain time and then continues indefinitely.
Fig.~\ref{initialcond}
illustrates the initial conditions for the case of the compact influx;
the case of the non-compact influx is similar.

The gauge freedom associated 
with the transformations $u \to f(u)$ and $v \to f(v)$ can be used to make $r$ 
linear in $u$ and $v$ on the initial null segments, so that
\be
r(u_0,v)=v \ , \qquad  r(u,v_0)=v_0+ u \, r_{,u}(u_0,v_0) \  . \label{icondr}
\ee
Defining $r_0 \equiv r(u_0,v_0)$,
equations~(\ref{icondr})
show that
\be
  r_0 \equiv v_0
  \ .
\ee
We choose
\be
r_0 = 5 \ ,
\ee
which is well outside the outer horizon.
This ensures that the null segment $u = u_0$
along which the initial influx~(\ref{pulse}) or (\ref{pulsenc})
of scalar field $\varphi$ is specified
is everywhere well outside the horizon.
The Misner-Sharpe \cite{Misner} mass function, the total effective mass inside a sphere of radius $r(u,v)$,
is equal to
\bq
m(u,v)&=&\frac{r}{2}\left(1+\frac{q^2}{r^2}-g_{rr}^{-1}\right)  \nn \\ 
&=&\frac{r}{2}\left(1+\frac{q^2}{r^2}+4\frac{r_{,u} r_{,v}}{2 e^{2\sigma}}\right)\label{massinf}\,.
\eq
Equation~(\ref{massinf}) implies that
$r_{,u}(u_0,v_0)$
in equation~(\ref{icondr})
is related to the initial 
mass $m_0$ and charge $q$ of the black hole by
\be
\label{ru0}
r_{,u}(u_0,v_0)=\frac{1}{4}\left[\frac{2}{r_0}\left(m_0-\frac{q^2}{2r_0}\right)-1\right]\,.
\ee

For the Brans-Dicke field $\Phi$, which is observationally constrained to vary little outside the black hole, we choose initial conditions 
\be
\Phi(u_0,v)=1=\Phi(u,v_0) \ .
\ee
This implies
\be \Phi_{,u}(u,v_0)=0=\Phi_{,uu}(u,v_0)\ , \ee
\be \Phi_{,v}(u_0,v)=0=\Phi_{,vv}(u_0,v) \ .\ee 
The constraint equations~(\ref{einuu}) and  
(\ref{einvv}) then become 
\bq
&&\sigma_{,u}(u,v_0) = \frac{r_0+u \, r_{,u}(u_0,v_0)}{2 r_{,u}(u_0,v_0)} 
[\varphi_{,u}(u,v_0)]^2\ , \label{const1} \\
&&\sigma_{,v}(u_0,v) =\frac{v}{2}[\varphi_{,v}(u_0,v)]^2\ \label{const2}.
\eq

We assume that there is no outflux along the initial null segment
$v = v_0$
\be
\varphi_{,u}(u,v_0)=0 \ . \label{ioutflux} 
\ee 
Equation~(\ref{ioutflux})
implies that the scalar field $\varphi$ is constant along the null segment $v=v_0$,
and, since $\varphi$ is defined only up to an overall constant,
without loss of generality we take this constant to be zero,
\be
\varphi(u,v_0)=0 \ . \label{iphi}
\ee
Equations~(\ref{const1}) and (\ref{ioutflux}) together imply that $\sigma_{,u} (u,v_0)=0$, which implies that $\sigma$ is constant along the null segment $v = v_0$.
The constant can be chosen arbitrarily,
since the choice~(\ref{icondr})
still leaves a gauge freedom in the overall scaling of $u$,
and we choose
\be \sigma(u,v_0)=\sigma(u_0,v_0)=-\ln(2)/2 \ , \ee
equivalent to choosing $g_{uv}(u_0,v_0) = 1$
in the line-element~(\ref{metric}).

The last initial condition to be specified is the influx
$\varphi_{,v}(u_0,v)$
of the scalar field along the null segment $u = u_0$.
We consider two cases,
one consisting of a compact pulse of influx,
the other a non-compact influx.
The compact pulse represents an accretion event
that takes place over a finite time,
while the non-compact influx represents accretion that
continues into the indefinite future.

For the compact pulse, we choose the influx of scalar field
along the initial null segment $u = u_0$ to be
\be
\varphi_{,v}(u_0,v)=A \sin^2\left(\pi \frac{v-v_0}{\Delta v}\right)
\label{pulse}\ , \ \ v_0\le v \le v_0+\Delta v \ ,
\ee 
vanishing outside the interval $v_0$ to $v_0 + \Delta v$.
We choose an interval
\be
  \Delta v = 1
  \ .
\ee 
The pulse has amplitude $A$.
For $A=0$, there is no pulse, and numerical integration yields the unperturbed RN black hole (see subsection \ref{unperturbed}).
The behavior of $\sigma$ along the null segment $u=u_0$
follows from integrating equation~\eqref{const2}.

For the non-compact influx, we choose the influx of scalar field
along the initial null segment $u = u_0$ to be
\be \varphi_{,v}(u_0,v)={A \over v} \ , \ \ v\ge v_0 \ . \label{pulsenc} \ee

\begin{figure}[th!]
\includegraphics[width=3.6in,keepaspectratio]{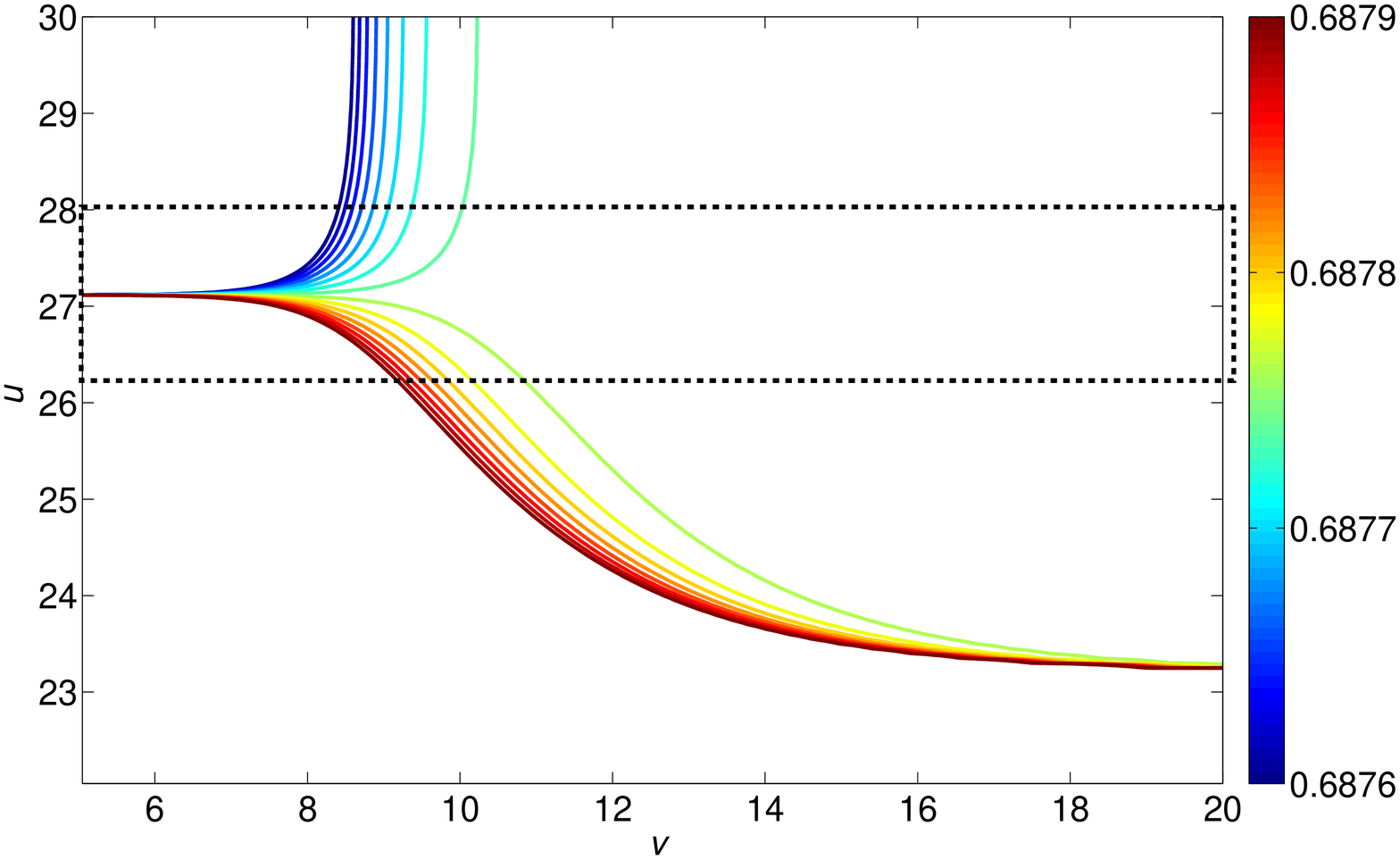}
\includegraphics[width=3.6in,keepaspectratio]{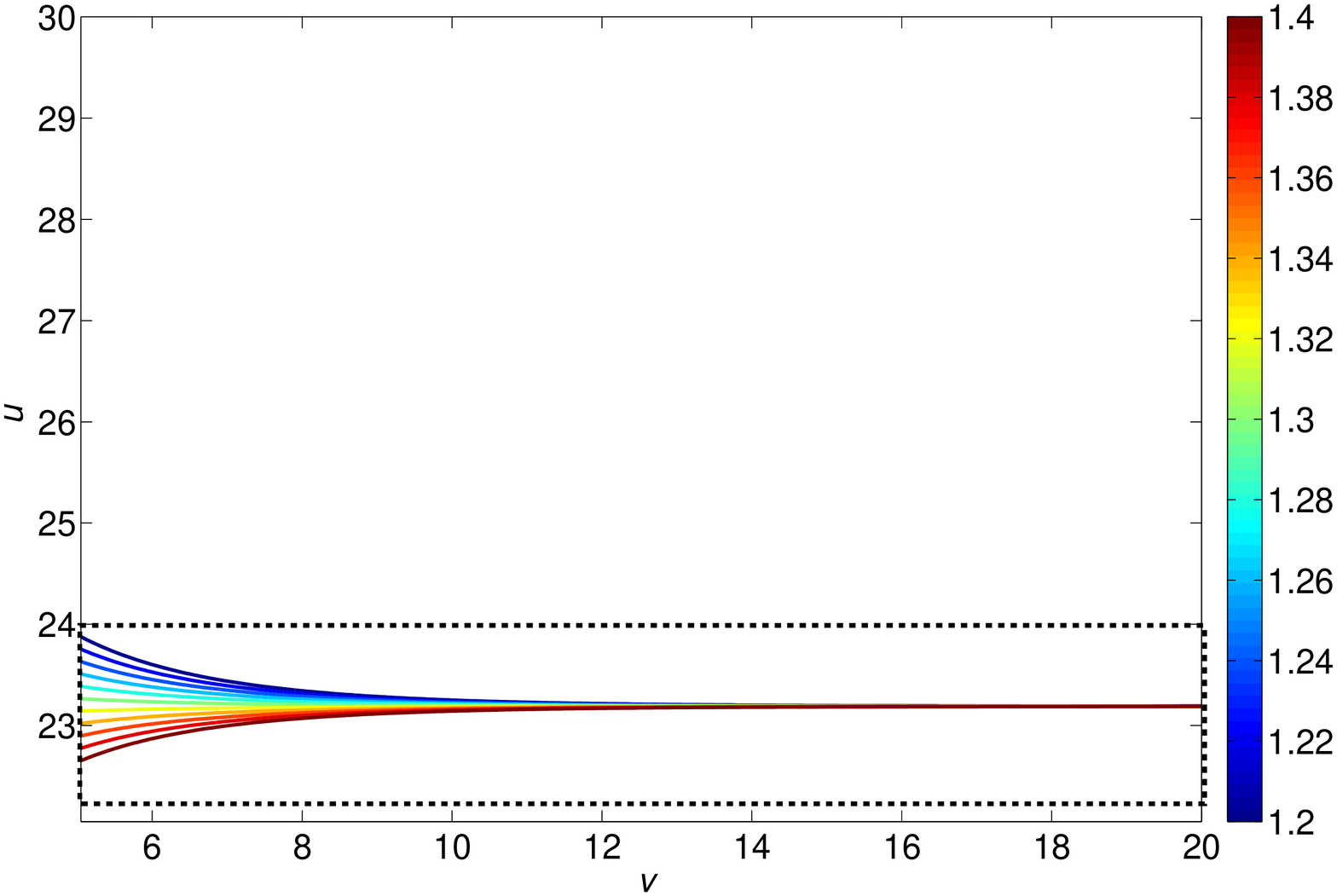}
\caption{\label{figure1}  The radial coordinate
$r$ as a function of $u$ and $v$ 
close to the inner ingoing apparent horizon (top) and outer ingoing apparent horizon (bottom),
for the unperturbed RN 
black hole with $m=1$ and $q=0.95$.
Horizons are at $r_- \approx 0.6878$ and $r_+ \approx 1.31$, in 
agreement with analytical expectation.
Contours are plotted only for radii in the range of the color bar,
$r \in [ 0.6876 , 0.6879 ]$ in the top panel,
$r \in [ 1.2 , 1.4 ]$ in the bottom panel.
}
\end{figure}

\begin{figure}[th!]
\begin{picture}(0,0)(0,0)
\put(190,180){$\mathcal{I}^+_R$}
\put(190,120){$\mathcal{I}^-_R$}
\put(15,180){$\mathcal{I}^+_L$}
\put(15,120){$\mathcal{I}^-_L$}
\put(120,242){$v=\infty$}
\put(87,200){$v=v_0$}
\put(158,160){$u=u_0$}
\put(116,147){$u$}
\put(130,147){$v$}
\put(161,250){$r=0$}
\put(30,250){$r=0$}
\put(172,97){$r=\mbox{constant}$}
\end{picture}
\includegraphics[width=2.85in,keepaspectratio]{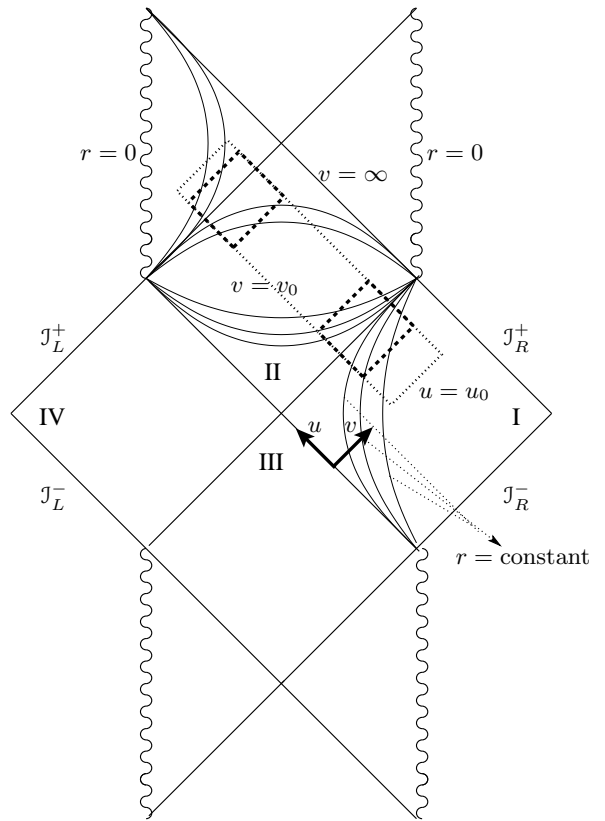}
\caption{\label{figurecp}  Carter-Penrose diagram for the eternal RN black hole. Some constant $r$ lines are depicted. The dotted line encloses a rectangular region in the $u,v$ plane, where the integration takes place (in the chosen coordinates $u_0=0\le u<30$, $v_0=5\le v <20$). The dashed lines enclose two regions corresponding to the regions enclosed by dashed lines seen in Fig.~\ref{figure1}. The same patterns of $r= \mbox{constant}$ lines are seen.}
\end{figure}

\subsection{\label{num} Numerical algorithm}

We integrate the field equations using a second order algorithm similar to 
that used by other authors (see for example \cite{Hansen:2005am} and 
references therein). The four evolution equations can be written as 
\be
\label{Xsolve}
Y^j_{,uv}=F^j(Y^k,Y^k_{,u},Y^k_{,v}) \ , \ k=1...4\ ,
\ee 
with $Y^j=r,\sigma,\Phi,\varphi$ for $j=1$ to $4$ respectively. 
The numerical code  determines the values of the variables $Y^j$ in the 
$(u,v)$ numerical grid solving, for each $v=v_0+n_v \Delta v$, 
equation~(\ref{Xsolve}) for $u=u_0+n_u\Delta u$ (starting at $n_u=1$) 
until the last point $(u_f,v_f)$ is finally reached. We performed standard 
numerical tests by running various simulations with different 
resolutions ($\Delta u$ and $\Delta v$), verifying that the code is indeed 
accurate to second order in $\Delta u$ and $\Delta v$. 

Although in double-null coordinates all the fields are regular,
there is a numerical difficulty across the event 
horizon of the black hole, since $r_{,u}$ diverges along the event  
horizon (see Fig.~\ref{figure1} -- bottom). In order to maintain numerical precision,
we used an algorithm with  
adaptive mesh refinement (AMR) in one direction. Given that strong gradients 
appear mainly in the $u$-direction, the AMR algorithm modifies the increment 
$\Delta u$ in such a way that $ \Delta r /r < 
10^{-4}$. This unidirectional approach is possible since Courant-like 
stability constraints on the step size are avoided by the use of null 
coordinates.

\subsection{\label{unperturbed} Unperturbed black hole}

Fundamental to the interior structure of a black hole is the location of its apparent horizons.
Fig.~\ref{figure1} reveals the location of these horizons in the $u,v$ plane for the unperturbed case. Fig.~\ref{figure1} shows the coordinate $r$ as a function of $u$ and $v$  for the unperturbed RN black hole (without accretion, $A = 0$). Recall that an apparent horizon may be loosely defined as the boundary between the region where outwardly directed light rays (i.e.\ with $u= \mbox{constant}$ and increasing $v$ in null coordinates) move outwards (i.e.\ to larger $r$) and the region where outwardly directed light rays move inwards. Apparent horizons are defined locally, and their locus is not, in general, a covariant statement; it depends on the chosen time foliation of the spacetime. An illustrative example is that the Schwarzschild solution may be foliated such that it has no apparent horizon \cite{Wald:91}, despite the presence of an event horizon, which is a covariant (and global) statement.

In terms of the chart (\ref{metric}), apparent horizons are the set of points $(u_a,v_a)$ such that
\be \frac{\partial r}{\partial v}(u_a,v_a)=0  \ . \label{ahorizon}\ee
This equation defines a curve in the $u,v$ plane.  Fig.~\ref{figure1} shows that
the apparent horizons lie along
$u_a= \mbox{constant}$,
and that $r(u_a,v_a)\simeq r_{\pm}$ to good accuracy,
meaning that the radii of the apparent horizons coincide
with the expected radii of the outer and inner event horizons, equations~\eqref{horizons}. This is a good test of the code. As a further test we were also able to reproduce the results reported by \cite{Hansen:2005am}.

The content of  Fig.~\ref{figure1} can also be understood in the context of the Carter-Penrose diagram of an eternal RN black hole. Fig.~\ref{figurecp} shows the Carter-Penrose diagram, outlining the regions covered by Fig.~\ref{figure1}. The diagram reveals the same pattern of lines of constant $r$ as Fig.~\ref{figure1}.

\section{\label{homog} Homogeneous approximation}
Results will be presented in section~\ref{resul},
but first it is useful to gain insight by applying the homogeneous approximation
\cite{Burko:1997xa,Burko:1998az,Burko:1998jz}.

If the black hole is accreting at a low rate,
it is reasonable to expect that the geometry,
at least outside the region where mass inflation is occurring,
would approximate that of a stationary black hole,
a Reissner-Nordstr\"om black hole.
A stationary black hole has the property of time translation invariance
\be
\label{homogeneousapproximation}
{\partial \over \partial t} = 0
\ .
\ee
Inside the horizon,
the time coordinate $t$ is spacelike, and
\cite{Burko:1997xa,Burko:1998az,Burko:1998jz}
has called the proposition that equation~(\ref{homogeneousapproximation})
holds inside the horizon the homogeneous approximation.
\cite{Hamilton:2008zz}
point out that the homogeneous approximation~(\ref{homogeneousapproximation})
fails in the inflationary zone even in the limit of infinitesimal accretion rates.
Rather, the homogeneous approximation~(\ref{homogeneousapproximation})
is equivalent to the assumption of symmetrically equal ingoing and outgoing fluxes.
The initial conditions prescribed in subsection~\ref{Iconditions} are asymmetric
between ingoing and outgoing,
so the homogeneous approximation~(\ref{homogeneousapproximation})
is not expected to hold.

However, the homogeneous approximation~(\ref{homogeneousapproximation})
simplifies the mathematics, because it means that all quantities are
functions only of the radial coordinate $r$.
Although the assumption of symmetrically equal ingoing and outgoing fluxes
is not realistic, nevertheless the behaviour in this case may provide
insight into the more realistic case of unequal ingoing and outgoing fluxes.

\subsection{Field equations}
\label{homofieldeqs}

The spherical symmetric, homogeneous line-element is given by eq.~\eqref{sphericalmetric}. The $tt$ and $rr$ components of the Einstein equations~(\ref{einstein}) are then
\bq
\frac{g_{rr}- {g_{rr}}^2 - rg_{rr}'}{r^2 {g_{rr}}^2} &=& 
\frac{8\pi}{\Phi} \left(^M T^{t}_{t} + ^{\Phi}  T^{t}_{t}\right)\label{gtt}\,,\\
\frac{g_{tt}-g_{tt} g_{rr}  + r g_{tt}'}{r^2 g_{tt}g_{rr}} &=& 
\frac{8\pi}{\Phi} \left(^M T^{r}_{r} + ^{\Phi} T^{r}_{r}\right)\label{grr}\,,
\eq
where the prime denotes a derivative with respect to $r$.
Energy-momenta in angular directions
follow from the energy-momenta in $t$ and $r$ directions,
together with the equations of covariant conservation of energy-momentum.
The non-vanishing
components of $^\varphi T_{\mu \nu}$, $^F T_{\mu \nu}$ and 
$^\Phi T_{\mu \nu}$ are,
in $\{ t , r , \theta , \phi \}$ coordinates,
\bq
^\varphi T^\mu_\nu &=& -\frac{(\varphi')^2}{8\pi g_{rr}}\, {\rm diag}(1,-1,1,1) \ ,  \\
^FT^\mu_\nu &=& -\frac{q^2}{8\pi r^4}\, {\rm diag}(1,1,-1,-1) \ , 
\\
^\Phi T^t_ t  &=& \frac{1}{8\pi g_{rr}}\left[\frac{\Phi'}{2}\left(\frac{g_{rr}'}{g_{rr}}
-\omega \frac{\Phi'}{\Phi}-\frac{4}{r}\right)-\Phi''\right]
\ ,  \\
^\Phi T^r_r  &=&  \frac{\Phi'}{16\pi g_{rr}}\left(-\frac{g_{tt}'}{g_{tt}}
+\omega \frac{\Phi'}{\Phi}-\frac{4}{r}\right)
\ ;  
\eq
whence
\be
\label{TM}
^M T= -\frac{(\varphi')^2}{4\pi g_{rr}}\,.
\ee
In the above we have taken into account that the black hole charge will 
give rise to an electrostatic field which is purely radial and consequently 
the only non-zero components of the electromagnetic tensor, $F_{\mu \nu}$ 
are $F_{t r} = - F_{r t}$. 
It is then straightforward to show from Maxwell's equations that
\be
\label{maxwell2}
F_{t r}= -\frac{ q \sqrt {-g_{tt} g_{rr}}}{r^2}\ .
\ee

The wave equation~(\ref{eqphif}) for the scalar field $\varphi$ can be written as a first order equation
\be
\varphi' \propto \sqrt{\left|\frac{g_{rr}}{g_{tt}}\right|}\,  \frac{1}{r^{2}} \ , \label{scalarhomo}
\ee
while the wave equation~(\ref{eqPhibd}) for the Brans-Dicke field $\Phi$
becomes
\bq
\Phi''&+& \frac{\Phi'}{2}\left(\frac{g_{tt}'}{g_{tt}}-\frac{g_{rr}'}{g_{rr}}+
\frac{4}{r}\right)
= g_{rr}
\frac{8\pi(^MT)}{3+2 \omega}\label{Phiwave}\ .
\eq
>From equations~\eqref{TM} and \eqref{scalarhomo}--\eqref{Phiwave}, one reads that the source term for the variations of $\Phi$ is proportional to
\be
\label{Phisource}
\frac{1}{3+2\omega}\frac{g_{rr}}{g_{tt}}\frac{1}{r^2} \ .
\ee
Thus, we conclude that the source term for $\Phi$ variations, depends strongly on the ratio $g_{rr}/g_{tt}$. We shall return to this point in section~\ref{resultsh}.

\subsection{\label{generalizedscalar} Generalized scalar field}

A massless scalar field has sound speed equal to the speed of light.
As originally envisaged by \cite{Poisson:1990eh},
mass inflation is expected to result from relativistic counter-streaming between
an influx and outflux of scalar field. We now introduce a generalized scalar field, which allows us to vary the speed of sound and hence have models with and without mass inflation. Our goal is to see the consequences, within the homogeneous approximation, for the variations of the Brans-Dicke scalar.

Consider a generalized scalar field $\varphi$ whose action is
\be
S_\varphi = \int {\sqrt {-g}} {\cal L}_\varphi (\varphi,X) d^4 x  \ , \label{matteraction}
\ee
where $X \equiv \varphi_{,\mu} \varphi^{,\mu}/2$. It follows that the energy-momentum tensor of the generalized scalar field is that of a perfect fluid
\cite{Babichev:2008dy}
\be
^\varphi T^{\mu}_{\nu} = (\rho + p) U^{\mu} U_{\nu} + p \delta^{\mu}_{\nu}\ ,
\ee
with 4-velocity
\be
\label{Unu}
U_{\nu} \equiv {\varphi_{,\nu} \over {\sqrt {-2X}}} \ , 
\ee
and proper energy density and pressure
$\rho=2X p_{,X}-p$ and $p={\cal L}_\varphi$.
If $\varphi_{,\nu}$ 
is timelike then the scalar field $\varphi$ is formally equivalent to a 
perfect fluid.
Generally, $\varphi_{,\nu}$ need not be timelike,
but it is timelike in the homogeneous case (cf. 
subsection~\ref{singlefluid}).

In the general case $p \equiv p(\varphi,X)$, the 
pressure cannot be expressed only in terms of $\rho$, since for a fixed 
$X$ and $p$ at a given event it will always be possible to have 
different $p_{,X}$ by changing the value of $\varphi$ at that event.
On the other hand, if $p \equiv p(X)$ then both $\rho$ and $p$ will be 
functions of $X$ alone, and consequently the relation between $\rho$ and 
$p$ is established once $X$ is fixed. If $p \propto X^{(1+w)/(2w)}$ with $w= {\rm constant}$,
then 
\be p=w \rho \ .
\ee
The corresponding sound speed $c_s$ is
\be
c_s = \sqrt{d p \over d \rho} = \sqrt{w} \ .
\ee
For example, the Lagrangian density 
${\cal L}_\varphi \propto X$ yields the usual 
massless scalar field model with equation of state $p=\rho$,
the model discussed in the previous subsection~\ref{homofieldeqs}.
As another example,
${\cal L}_\varphi \propto X^2$ describes an ultra-relativistic fluid with 
equation of state $p=\rho/3$.
As a final example,
${\cal L}_\varphi \propto X^n$ as $n \rightarrow \infty$
describes a pressureless fluid (dust).

\subsection{Homogeneous generalized scalar field}
\label{singlefluid}

The homogeneous approximation implies that $\varphi$ is
a function of $r$ only.
The only non-vanishing component of the 4-velocity
$U^\nu \propto U_\nu \propto \varphi_{,\nu}$,
equation~(\ref{Unu}),
is therefore $U^r$.
Since $g_{rr}<0$ inside the horizon,
$U^r$ is
\be
U^r=-\frac{1}{U_r}=\mp \sqrt{|g^{rr}|} \ , \ee
where the $\mp$ sign is $-$
for $\varphi'>0$ and $+$ for $\varphi'<0$.
The 4-velocity $U^\nu$ is timelike, and therefore the behaviour of the generalized
scalar field is identical to that of a perfect fluid.
The energy-momentum $^\varphi T^{\mu}_{\nu}$ of the generalized scalar field is
\be
{}^\varphi T^{t}_{t}= {}^\varphi T^{\theta}_{\theta}= {}^\varphi T^{\phi}_{\phi} = p = w \rho = -w {}^\varphi T^{r}_{r}\, , 
\ee
and the trace of the matter energy-momentum tensor becomes
\be
{}^M T=\rho(3w-1) \ . \label{trace} \ee
The proper energy density $\rho$ in terms of the metric coefficients follows from energy-momentum conservation, ${T^{\mu \nu}}_{;\nu}=0$, which implies 
\be
\rho= - T_{r}^{r}=\rho_i\left(\frac{g_{tti}}{g_{tt}}\right)^{(1+w)/2} 
\left(\frac{r_i}{r}\right)^{2(1+w)}\label{EMtensor}\ , 
\ee
where $\rho_i$, $g_{tti}$ and $r_i$ are all integration constants; $\rho_i$ is the density at the surface $r=r_i$, where $g_{tt}$ is $g_{tti}$. 

\subsection{Initial conditions}

Equations~\eqref{trace} and \eqref{EMtensor} define the right hand side~(\ref{Phisource}) of the BD scalar equation of motion, and will be used in subsection \ref{resultsh}. In solving the latter equation, we shall work in the limit where $\rho_i$ is small, and impose initial conditions inside the unperturbed outer horizon,
since the homogeneous approximation, which requires symmetrically equal
influx and outflux, fails outside the outer horizon.
Moreover, we take the initial conditions in equation~(\ref{EMtensor}) to be
 \be r_i=0.95 r_+ \ , \qquad -g_{tti}=\frac{1}{g_{rri}}=1-\frac{2m_0}{r_i}+\frac{q^2}{r_i^2} \ . \ee

\subsection{Results for the homogeneous approximation}
\label{resultsh}
We now present results for the homogeneous approximation
applied to a generalized scalar field inside a spherical charged black hole
in Brans-Dicke theory.
We consider two cases,
one with $w = 0$, the other with $w = 1$.

In the case with $w=0$, mass inflation does not occur and large variations of the Brans-Dicke scalar $\Phi$ are present. By contrast, in the $w=1$ case, mass inflation occurs and $\Phi$ variations are small. We will argue that it is actually mass inflation that prevents large $\Phi$ 
variations in the $w=1$ case. 

Within the homogeneous approximation, the task is to obtain the behaviour of the metric coefficients  in \eqref{sphericalmetric}, i.e.\ $g_{rr}$, $g_{tt}$, by solving (\ref{gtt}),  (\ref{grr}), and of the Brans-Dicke scalar field $\Phi$, by solving (\ref{Phiwave}). The matter scalar field $\varphi$ is given by equation~(\ref{scalarhomo}).

\begin{figure}
\includegraphics[width=3.7in,keepaspectratio]{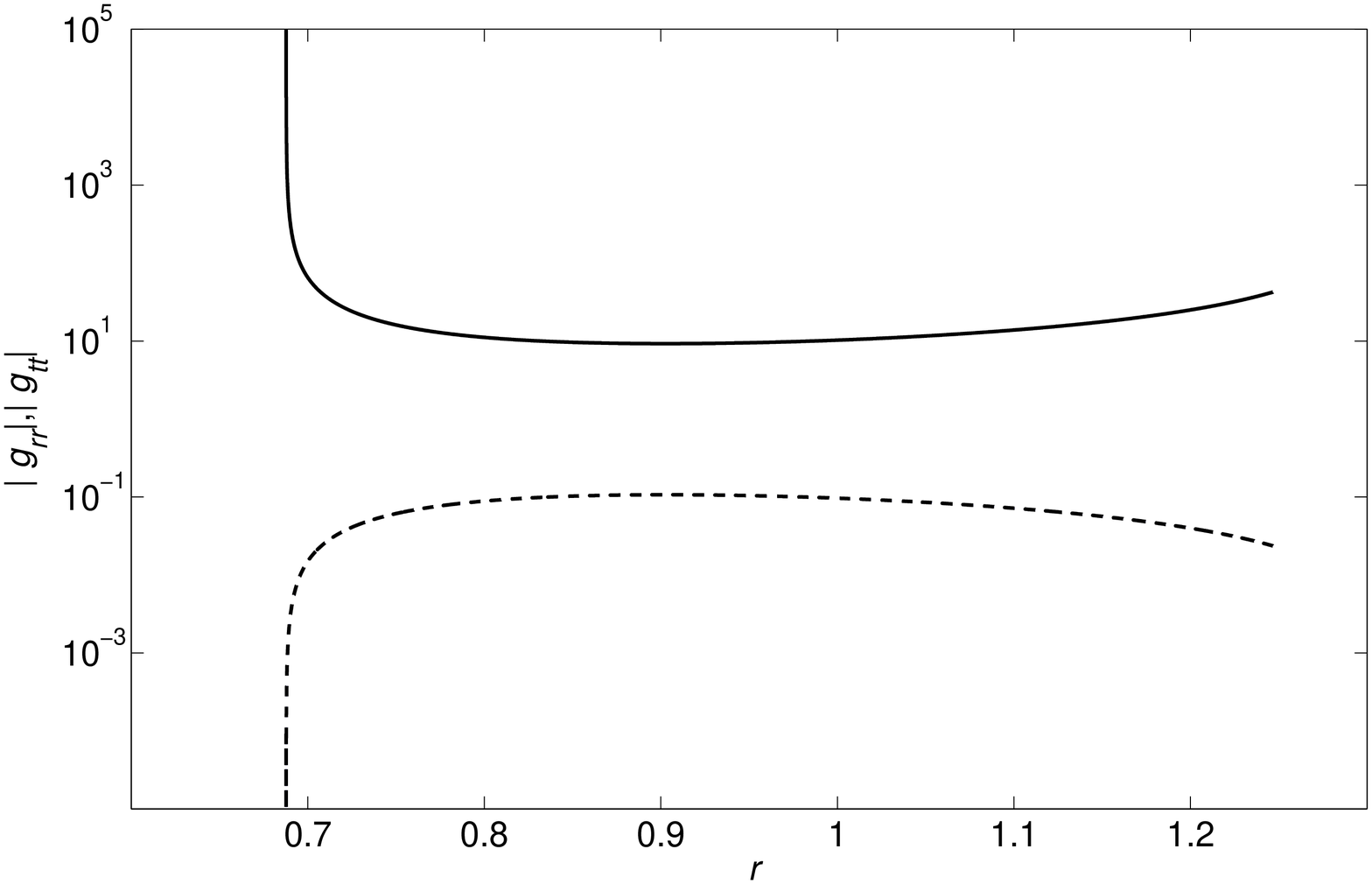}
\includegraphics[width=3.7in,keepaspectratio]{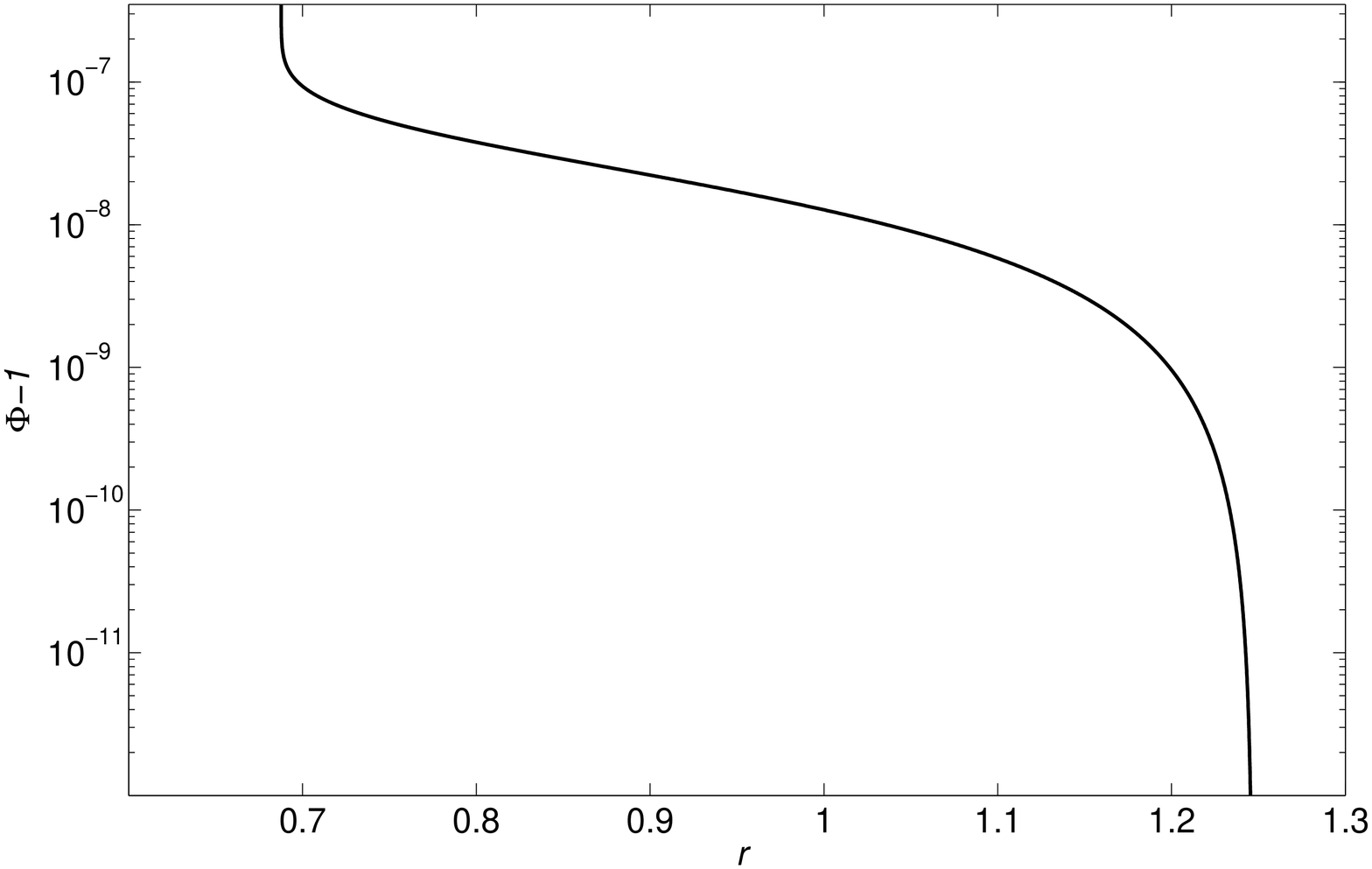}
\caption{\label{figure3}  The evolution of $|g_{rr}|$, $|g_{tt}|$ 
(top, solid and dashed lines respectively) and $\Phi-1$ (bottom) with $r$, in the case $w=0$ with $\rho_i=10^{-4}$. This behaviour of $|g_{rr}|$, $|g_{tt}|$ indicates the presence of a horizon around $r\simeq 0.69$. The evolution of $\Phi-1$ indicates that the BD scalar diverges as the horizon is approached.}
\end{figure}

Fig.~\ref{figure3} shows the result for the numerical integration of $g_{rr}$, $g_{tt}$ and $\Phi$, taking $w=0$.  The result is interpreted as follows. Equation (\ref{massinf}) shows that the interior mass is large if $|g_{rr}|$ is small, so a signature of mass inflation is that $|g_{rr}|$ is becoming exponentially small. This is not the case here. Rather,  Fig.~\ref{figure3} (top) shows that $|g_{rr}|$ diverges and $|g_{tt}|$ vanishes, around
$r\simeq 0.69$. This is characteristic not of mass inflation, but rather of matter hitting the inner horizon. Fig.~\ref{figure3} (bottom) indicates that the Brans-Dicke scalar is diverging as this horizon is approached. This is the expected behaviour of the BD scalar near a horizon, given a small variation of this scalar, as discussed in the introduction, cf. equation~\eqref{propintro}. It should be noted that this expectation relies on the homogeneous approximation, which breaks down at the outer horizon and outside.

\begin{figure}
\includegraphics[width=3.7in,keepaspectratio]{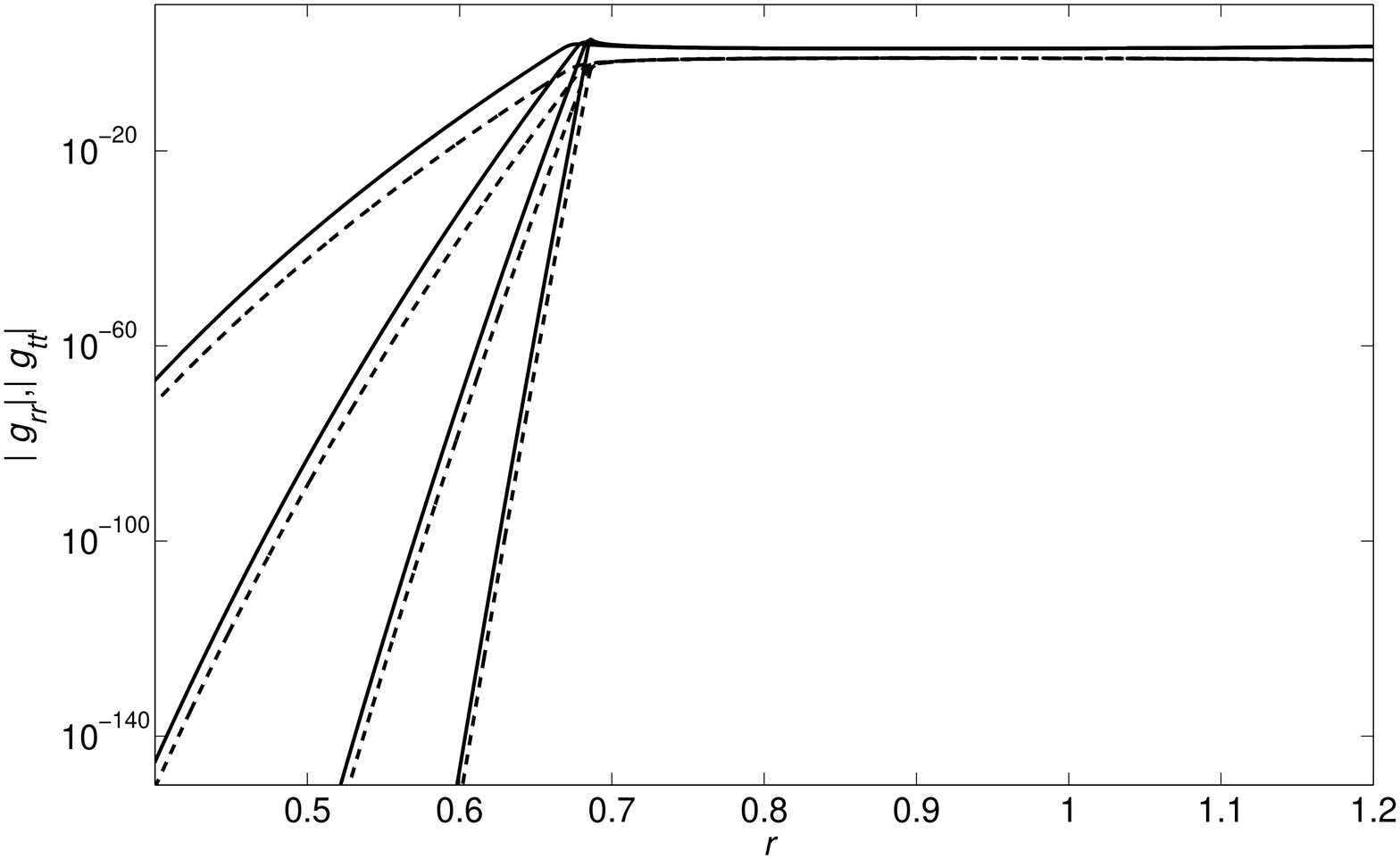}
\includegraphics[width=3.7in,keepaspectratio]{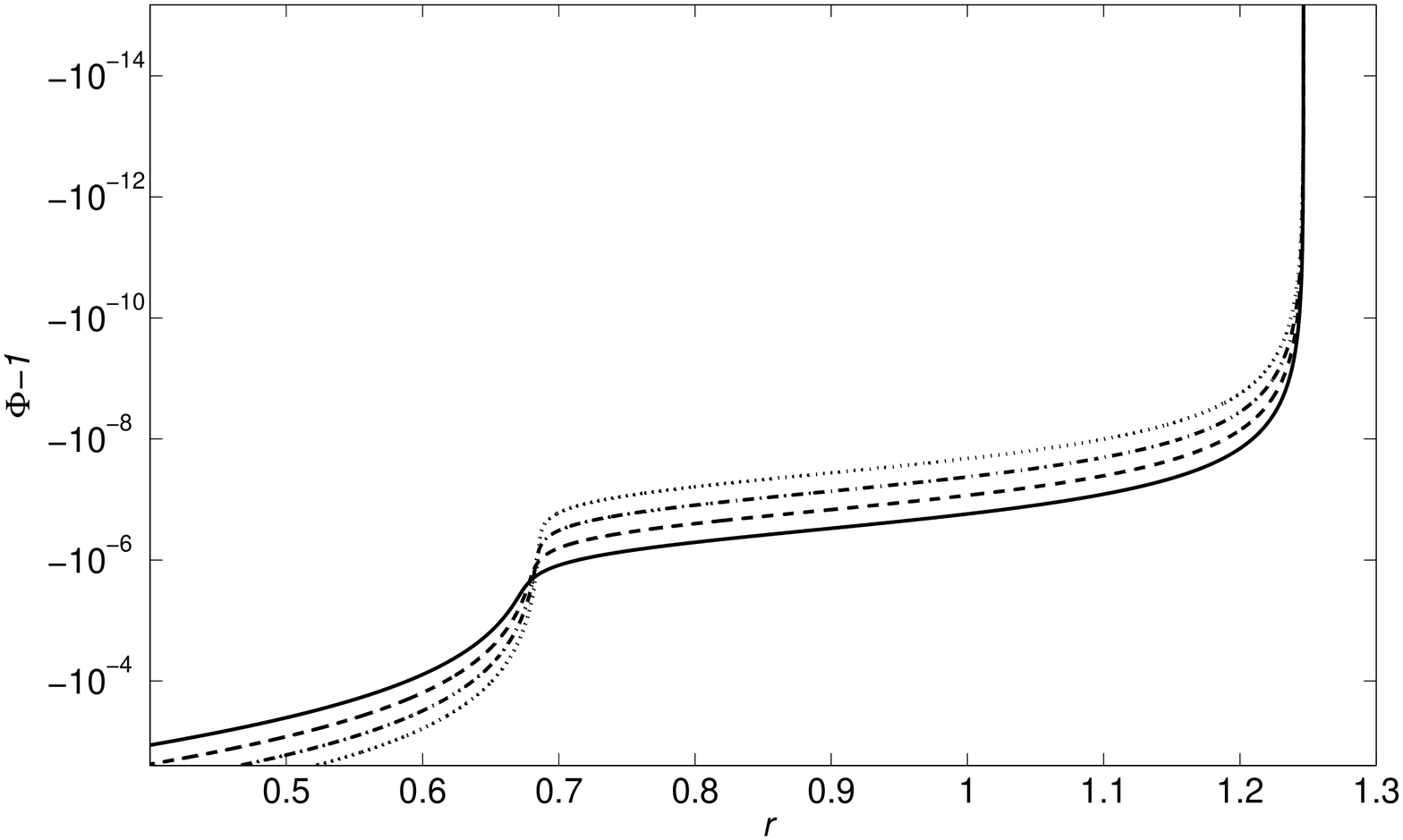}
\caption{\label{figure4}  The evolution of $|g_{rr}|$,$|g_{tt}|$ 
(top, solid and dashed lines respectively) and $\Phi-1$ (bottom) in the $w=1$ case for
various choices of $\rho_i$ (from right to left in top figure and from dotted to solid line in bottom figure, $\rho_i=10^{-4},2 \times 10^{-4},4 \times 10^{-4}, 8 \times 10^{-4}$). This behaviour of  $|g_{rr}|$, $|g_{tt}|$ indicates the existence of mass inflation.  A noticeable jump in $\Phi$ occurs at the beginning of mass inflation, but it is followed by a gentle variation of $\Phi$.}
\end{figure}

Fig.~\ref{figure4} shows the result for $g_{rr}$, $g_{tt}$ and $\Phi$, taking $w=1$.
This corresponds to the standard case of a massless scalar field,
the same as considered in section~\ref{model},
and can therefore be used to anticipate the results of section~\ref{resultsfull}.
The behaviour is quite different from the $w=0$ case. Fig.~\ref{figure4} (top) shows that \textit{both} $g_{tt}$ and $g_{rr}$ decrease sharply (and proportionally) around $r\simeq 0.69$. This indicates the presence of mass inflation. A counter-intuitive fact shown in this plot is that for smaller 
values of the initial energy density $\rho_i$, mass inflation is more 
abrupt. 

Fig.~\ref{figure4} (bottom) shows the behaviour of the Brans-Dicke scalar $\Phi$. Around $r\simeq 0.69$, where mass inflation starts, there is a clear jump in the variation of $\Phi$, which is followed by a gentle decrease in $\Phi$ to smaller radius. This indicates that the variations of $\Phi$ are quite small in the mass inflation region.

\begin{figure}
\includegraphics[width=3.7in,keepaspectratio]{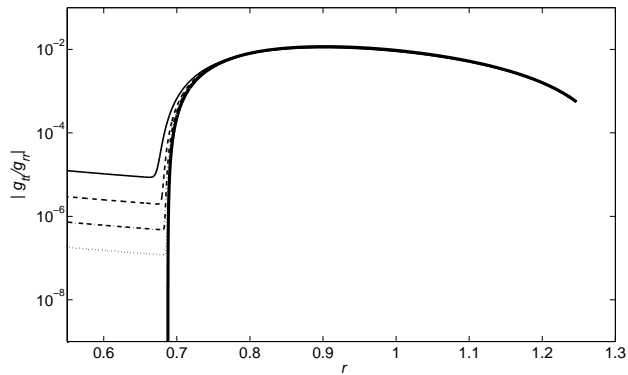}
\caption{\label{figureratio}  The evolution of $|g_{tt}|/|g_{rr}|$ for $w=0$ (dark solid line) and for $w=1$ (same cases as exhibited in Fig.~\ref{figure4}.)}
\end{figure}

Having described the behaviour obtained from numerical integration we now try to understand it analytically. The case of $w=1$, where mass inflation occurs, indicates that the variations of $\Phi$ are small (less than one percent); thus $\Phi \approx 1$.
It follows that $|^\Phi T_{\mu \nu}| \ll | ^M T_{\mu \nu}|$,
and consequently $^\Phi T^\mu_\nu$ may be neglected in comparison to $^M T_{\mu \nu}$ in equations~(\ref{gtt}) and (\ref{grr}).
In this approximation, the problem reduces to that of general relativity coupled to matter and a scalar field, without back-reaction from the BD scalar.  Moreover, Fig.~\ref{figure4} shows that the variations of $g_{rr}$ and $g_{tt}$ are quite fast in the mass inflation region. Thus, eqs. (\ref{gtt}) and (\ref{grr}) may be approximated by
\bq
\frac{g_{rr}'}{r{g_{rr}}^2} &\simeq& 
- 8\pi \, {}^M T^{t}_{t}=8 \pi w\, {}^M T^{r}_{r} \ , 
\label{betamassinfr}
 \\
\frac{g_{tt}'}{r g_{tt} g_{rr}}&\simeq& 
8\pi \, {}^M T^{r}_{r}\, 
\label{betamassinft}\, .
\eq
These two equations imply that
\be
\frac{g_{rr}'}{g_{rr}} \simeq   w\frac{g_{tt}'}{g_{tt}}\ \ \ \ \Rightarrow \ \ \ -g_{rr} \propto (g_{tt})^w\ . \label{prophomo}\ee 
Fig.~\ref{figureratio} shows the ratio $|g_{tt}|/|g_{rr}|$ for the scenario with $w=0$, where mass inflation does not occur, and the one with $w=1$, where it does. For the latter, the ratio is approximately constant in the mass inflation region, in accordance with \eqref{prophomo}, even though each metric coefficient individually varies rapidly, as shown in Fig.~\ref{figure4}.

Equation \eqref{prophomo}, for $w=1$, indicates that mass inflation occurs. It follows from (\ref{EMtensor}), (\ref{betamassinft}) and (\ref{prophomo}) that
\be
\frac{g_{rr}'}{g_{rr}} \propto \frac{1}{r^3}
\label{betaomega}\,,
\ee
in the mass inflation region. Then
\be
g_{rr} \propto \exp\left[-\left(\frac{\Delta r}{r}\right)^2\right]\,.
\ee
where $\Delta r$ scales roughly with ${\sqrt {\rho_i}}$. Thus, the larger $\rho_i$, the larger the scale $\Delta r$ for variations of $g_{rr}$. This agrees with the behaviour observed in Fig.~\ref{figure4} (top): mass inflation is more abrupt for smaller $\rho_i$.


Similarly, within this approximation scheme, we can obtain some analytical information for the variation of the Brans-Dicke scalar $\Phi$ with $r$,
using eq.~(\ref{Phiwave}). 
This equation may be written as 
\be
\label{Phihom1}
\Phi''+\frac{\Phi'}{2}\left[\left(\ln\frac{|g_{tt}|}{|g_{rr}|}\right)'+\frac{4}{r}\right]= \frac{g_{rr} }{(g_{tt})^{(1+w)/2}} f(r) \ ,
\ee 
where $f(r)$ is some function of $r$, whose form is unimportant for this argument. If mass inflation does not occur, as in the $w=0$ case, both the coefficient of $\Phi'$ and the source term diverge, from the behaviours of the metric coefficients seen in Fig.~\ref{figure3}. This sources a divergent behaviour of $\Phi$, as seen in Fig.~\ref{figure3}. If mass 
inflation occurs, as in the $w=1$ case, then equation~\eqref{prophomo} allows equation~(\ref{Phihom1}) to be written
\be
\label{Phihom2}
\Phi''+\frac{\Phi'}{2}\left[\left(\ln\frac{|g_{tt}|^{(1-w)}}{C}\right)'+\frac{4}{r}\right]= C(g_{tt})^{(w-1)/2} f(r) \ ,
\ee
where $C$ is a constant. Hence $w=1$ is a special case where the coefficient of $\Phi'$ and the source term become bounded as $g_{tt}\rightarrow 0$, and therefore the variations of $\Phi$ remain small in the mass inflation region.
This is consistent with the behaviour observed in Fig.~\ref{figure4} (bottom). Thus, one concludes that, for $w=1$, it is the \textit{existence of mass inflation that prevents large $\Phi$ variations} in the region around $r\simeq 0.69$.

To summarise, the homogeneous approximation indicates that for $w=1$, mass inflation should occur and that the induced variations of the BD scalar (and therefore of the coupling constant) within the mass inflation region are small.
In the next section,
we confirm these indications by a full non-linear numerical evolution of the model of section \ref{model}.

\section{\label{resul}Results}

In this section we present results for a massless scalar field $\varphi$ accreting
on to a spherical charge black hole in Brans-Dicke theory.
We consider two cases,
addressed in turn in subsections~\ref{resultsfull} and \ref{resultsnoncompact},
one in which the initial influx of scalar field is turned on for only a finite time,
equation~(\ref{pulse}),
and another in which the initial influx starts at some time
and then persists into the indefinite future,
equation~(\ref{pulsenc}).

The equations to be solved are the four
equations given by (\ref{scalar}) and \eqref{einuv}-\eqref{bdscalar},
which constitute a complete set of equations.
The remaining Einstein equations, eq.~(\ref{einuu}) and (\ref{einvv}),
are automatically satisfied by covariant conservation of energy-momentum,
and can be used to check the accuracy of the numerical integration.

\subsection{Full non-linear numerical evolution for a compact influx}
\label{resultsfull}
\begin{figure}
\includegraphics[width=3.6in,keepaspectratio]{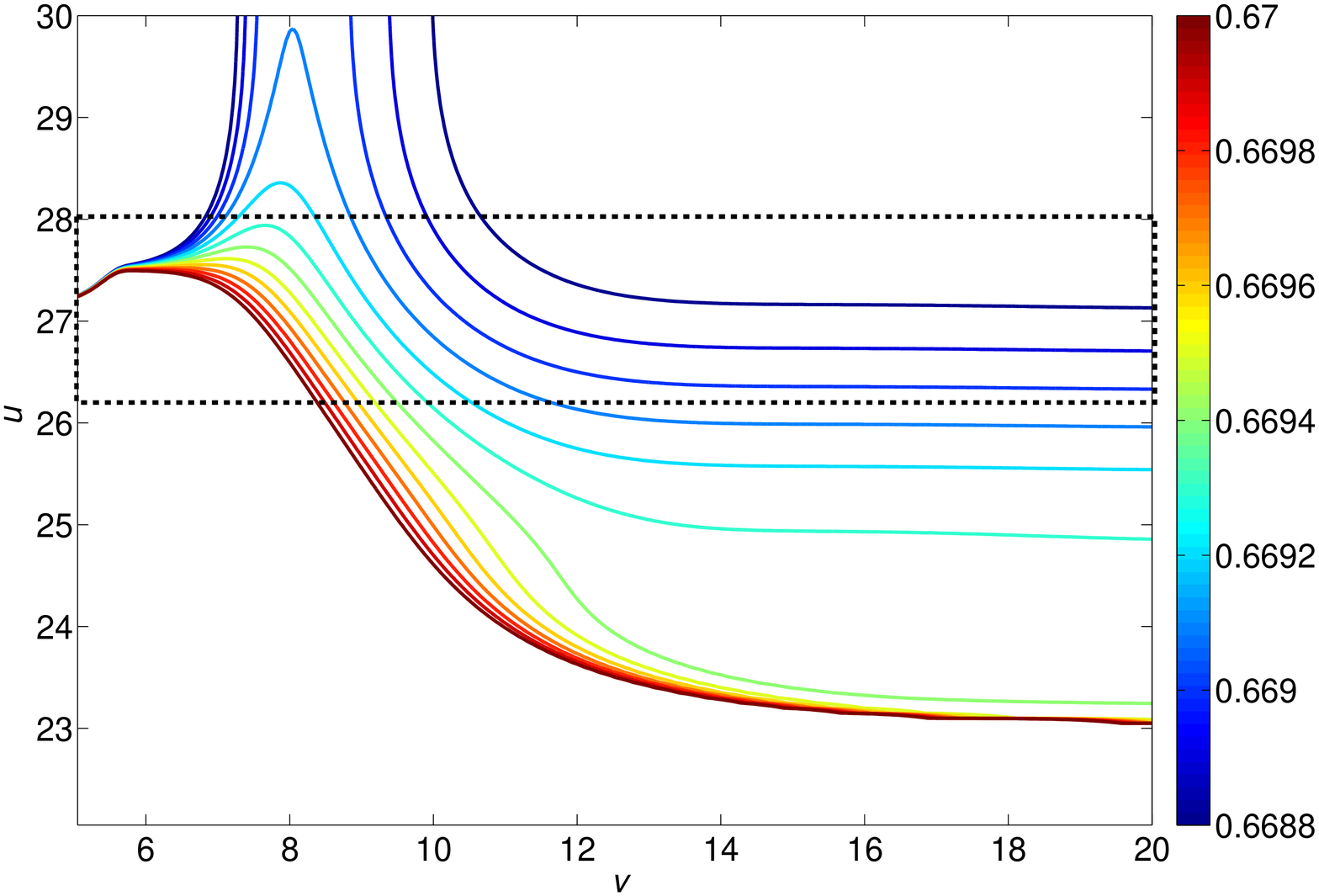}
\includegraphics[width=3.6in,keepaspectratio]{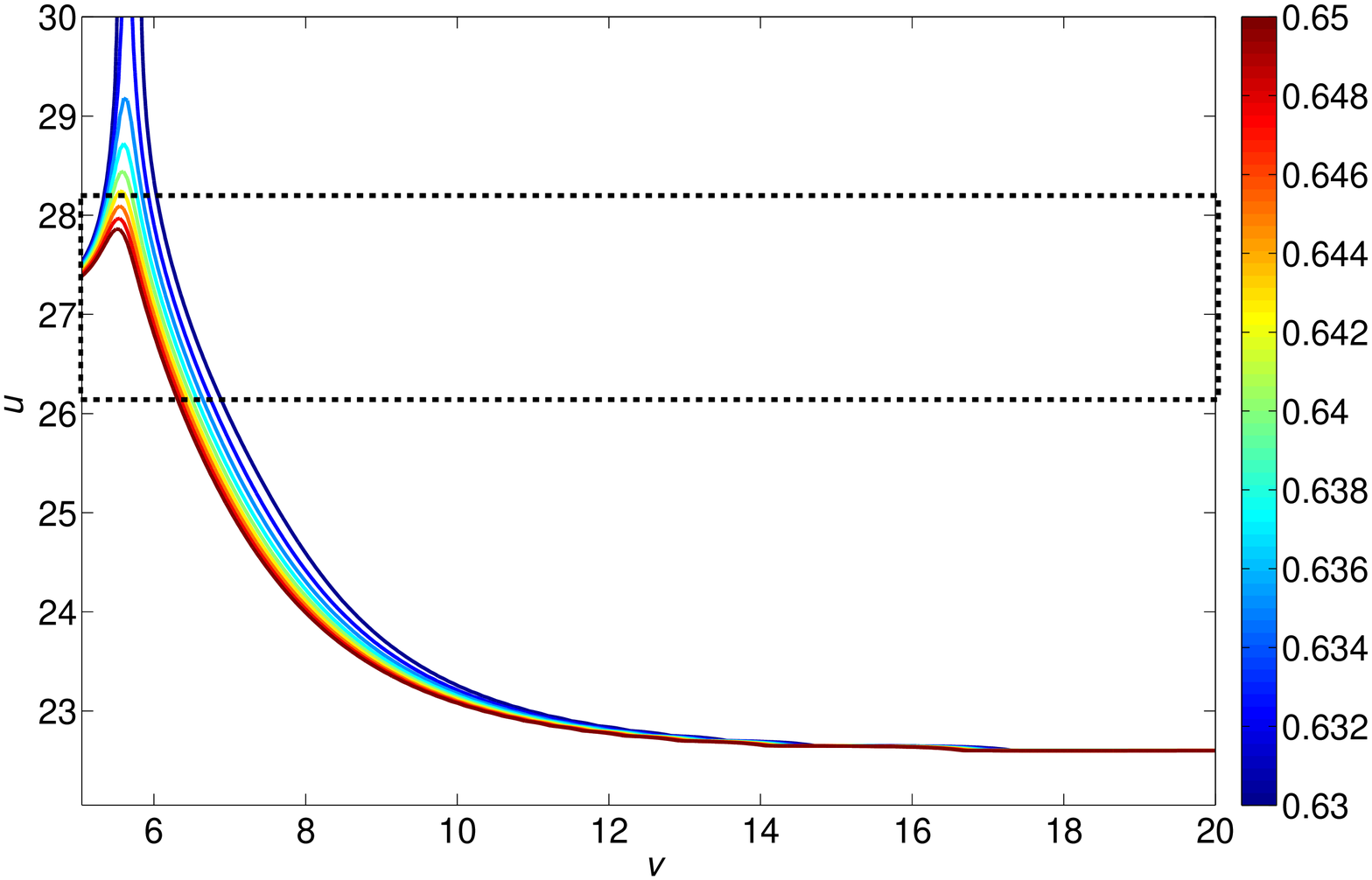}
\caption{\label{figure6b}   The radial coordinate
$r$ as a function of the null coordinates $u$ and $v$ close to the inner apparent 
horizon when a single compact pulse with amplitude $A=0.05$ (top) or $A=0.1$ (bottom)
and during $\Delta v=1$ is fed on to an initially unperturbed RN 
black hole with $m_0=1$ and $q=0.95$. The radius $r$ is no longer a 
monotonic function of $u$ as in the unperturbed case (Fig.~\ref{figure1} -- top). The locus of the inner apparent horizon is a curve that interpolates between the maxima
$\partial r / \partial v = 0$
of the $r=\mbox{constant}$ curves. For the pulse with larger amplitude, the inner apparent horizon changes more abruptly with $v$.}
\end{figure}

\begin{figure}
\includegraphics[width=3.6in,keepaspectratio]{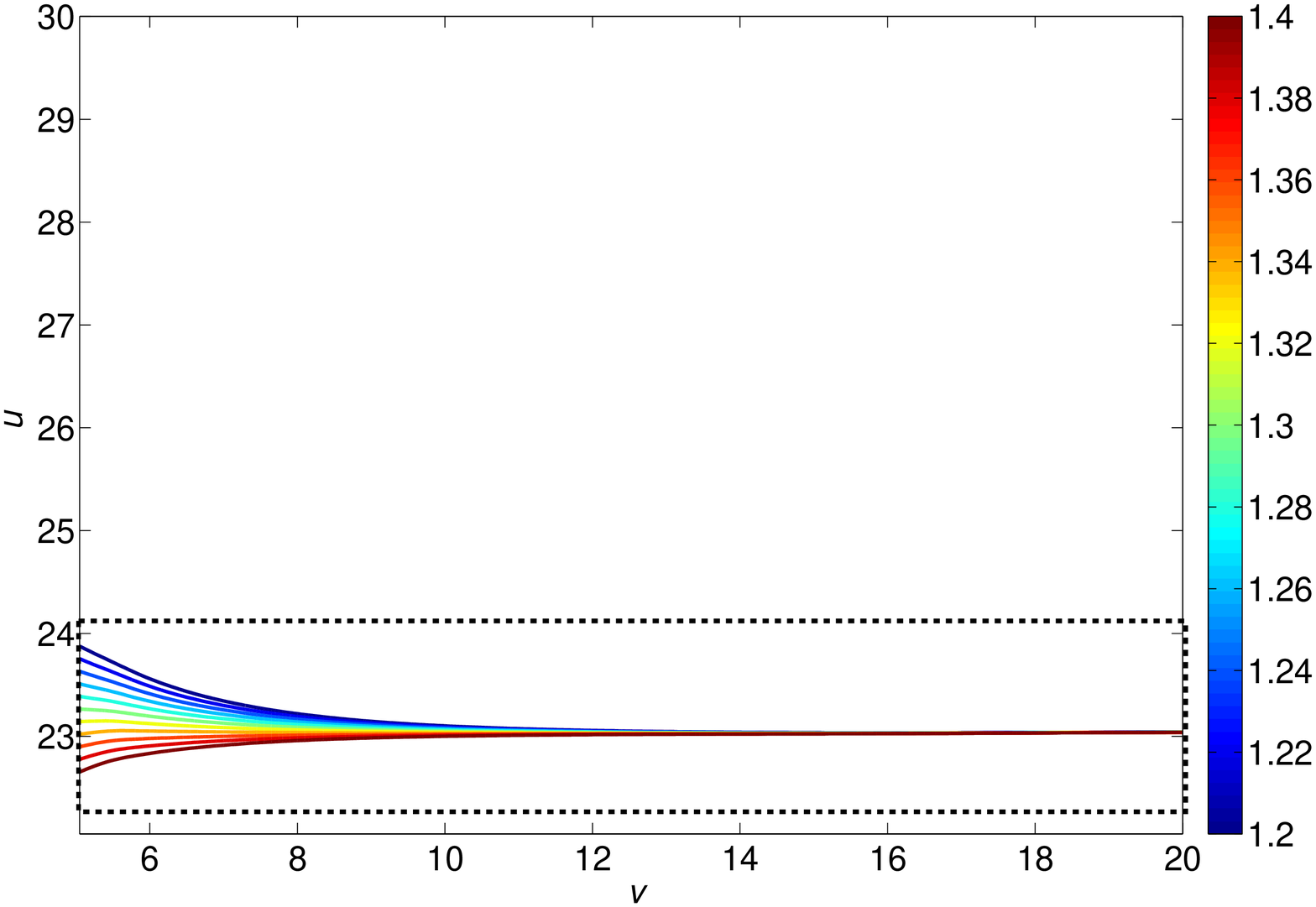}
\includegraphics[width=3.6in,keepaspectratio]{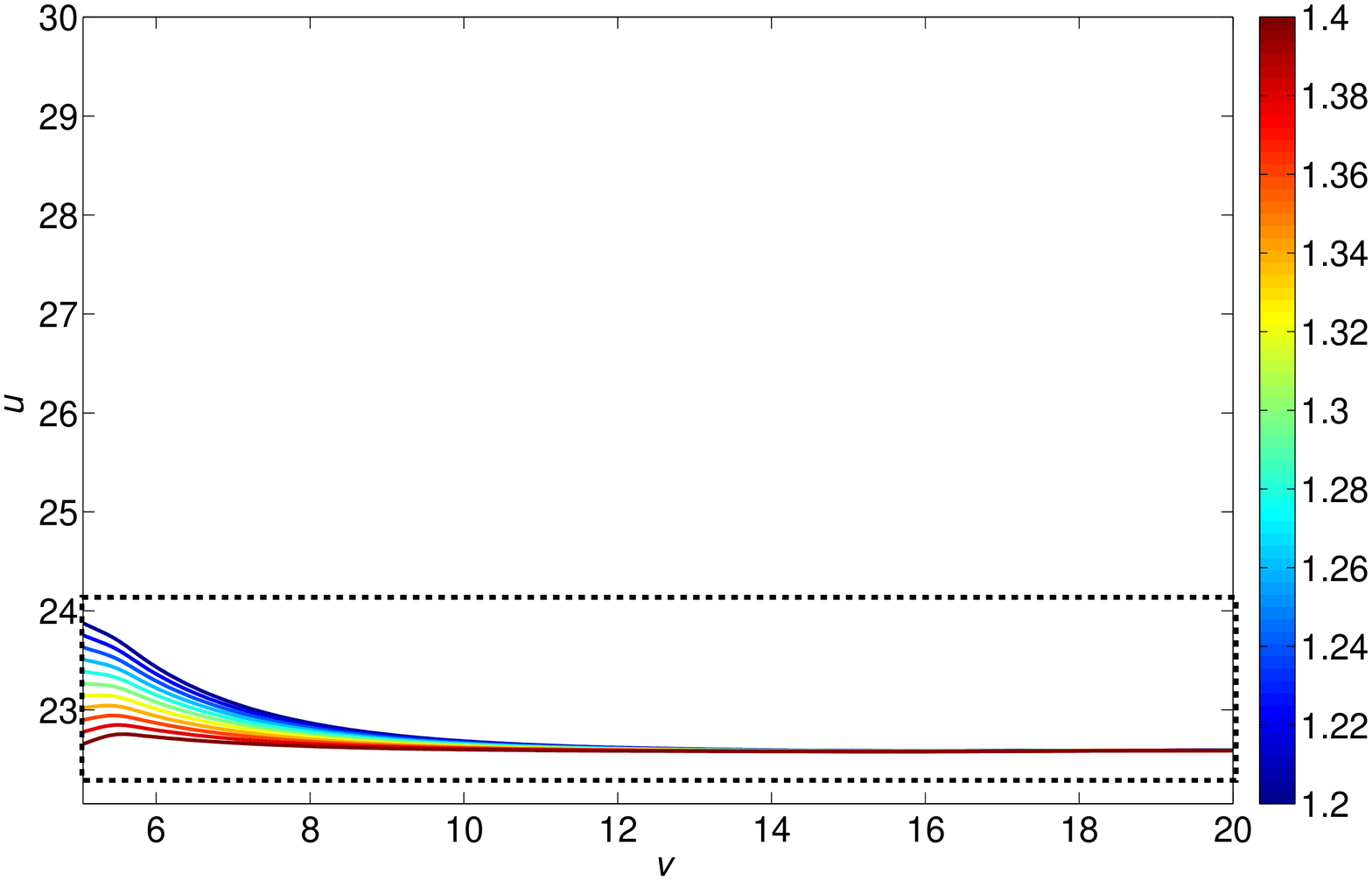}
\caption{\label{figure7b} This 
plot shows a detail of the radius $r$ in the vinicity of the outer horizon for the pulse with amplitude $A=0.05$ (top) or $A=0.1$ (bottom). The behaviour is similar to the case 
without accretion $A = 0$ (Fig.~\ref{figure1} -- bottom), which can be attributed to the fact that the accreted mass is small. The value of $u$ at large $v$ (in the future)
differs noticeably from the zero accretion case.}
\end{figure}

In this subsection,
the black hole is fed with a compact pulse of massless scalar field,
equation~(\ref{pulse}).

Fig.~\ref{figure6b} shows the value of the radial 
coordinate $r$ as a function of $u$ and $v$ close to the inner apparent 
horizon when a single compact pulse with amplitude $A=0.05$ or $A=0.1$, and 
duration $\Delta v=1$, is fed on to an initially unperturbed Reissner-Nordstr\"om 
black hole with $m=1$ and $q=0.95$. There are substantial 
changes compared to the Reissner-Nordstr\"om solution shown in Fig.~\ref{figure1}. For example, 
$r$ is no longer a monotonic function of $u$. In particular, the variation seen in Fig.~\ref{figure6b} implies, according to the definition in section \ref{num}, that the locus of the inner apparent horizon in the $u,v$ plane is not given by $u_a=\mbox{constant}$; rather $u_a$ increases with $v_a$. According to equation~(\ref{ahorizon}), the locus of the inner apparent horizon is described by a curve that interpolates between the maxima of the constant $r$ curves in Fig.~\ref{figure6b}. The value of $r$ on these maxima reveals that as time increases ($u$ and $v$ increase), the $r$ coordinate of the inner apparent horizon \textit{decreases}. This is to be expected. Indeed, the radii of the outer and inner horizons of a RN black hole vary with mass as
\be
\label{horizonr}
\frac{\partial r_{\pm}}{\partial m}= 1 \pm \frac{m}{\sqrt {m^2-q^2}}\, .
\ee
Consequently, increasing the mass of a Reissner-Nordstr\"om black hole 
with a fixed $q$ leads to a larger $r_+$ and a smaller $r_-$. Of course,
the geometry of an accreting black hole is not the same as the RN solution,
but the observed decrease in the radius of the inner apparent horizon
is consistent with equation~(\ref{horizonr}).

\begin{figure}[t!]
\begin{picture}(0,0)(0,0)
\put(190,180){$\mathcal{I}^+_R$}
\put(190,120){$\mathcal{I}^-_R$}
\put(15,180){$\mathcal{I}^+_L$}
\put(15,120){$\mathcal{I}^-_L$}
\put(142,224){$v=\infty$}
\put(87,200){$v=v_0$}
\put(158,160){$u=u_0$}
\put(116,149){$u$}
\put(130,149){$v$}
\put(30,250){$r=0$}
\end{picture}
\includegraphics[width=3.0in,keepaspectratio]{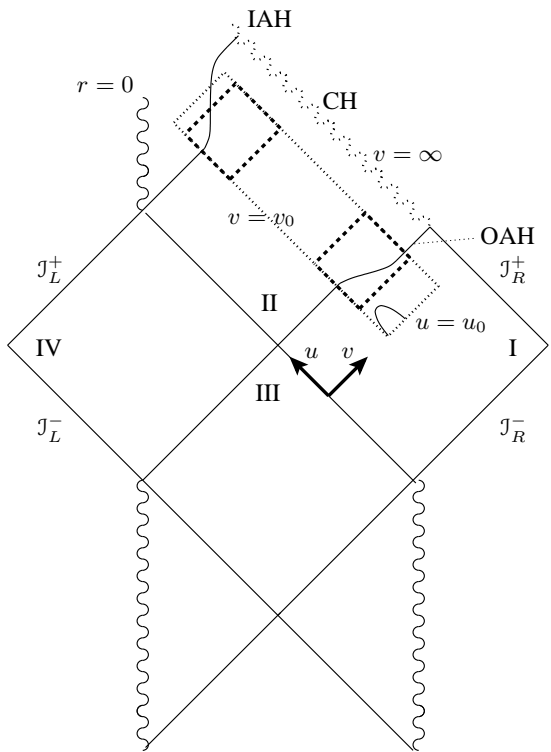}
\caption{\label{figurecpaccretion}  Heuristic Carter-Penrose diagram for the perturbed RN black hole. The initial perturbation is the curve depicted at the beginning of the integration box. The diagram is drawn to include the features observed in Fig.~\ref{figure6b} and \ref{figure7b}, namely that $u_a$ increases/decreases with $v_a$ for the inner apparent horizon (IAH)/outer apparent horizon (OUH). One expects the curvature to blow up at the Cauchy horizon (CH), and to become Planckian on a spacelike surface before the CH.}
\end{figure}

By the same token, accretion is expected to cause the outer apparent horizon to become slightly larger compared to the unperturbed RN solution. This expectation is confirmed in Fig.~\ref{figure7b}, which shows a detail around the outer apparent horizon for the same simulations as Fig.~\ref{figure6b}. Fig.~\ref{figure7b} shows two relevant features. Firstly, whereas at early $v$ times the plot is quite similar to that of the unperturbed case (Fig.~\ref{figure1} -- bottom), at late $v$ times there is a slight but clear decrease in the $u_a$ coordinate of the outer apparent horizon (clearly seen in Fig.~\ref{figure7b} -- bottom, but already noticeable in Fig.~\ref{figure7b} -- top). Secondly, $r(u_a,v_a)$ is at late $v$ times slightly, but again clearly, larger than the unperturbed value $1.31$, as expected. The behaviour just described can be summarised in the heuristic Carter Penrose diagram of Fig.~\ref{figurecpaccretion}.

\begin{figure}
\includegraphics[width=3.7in,keepaspectratio]{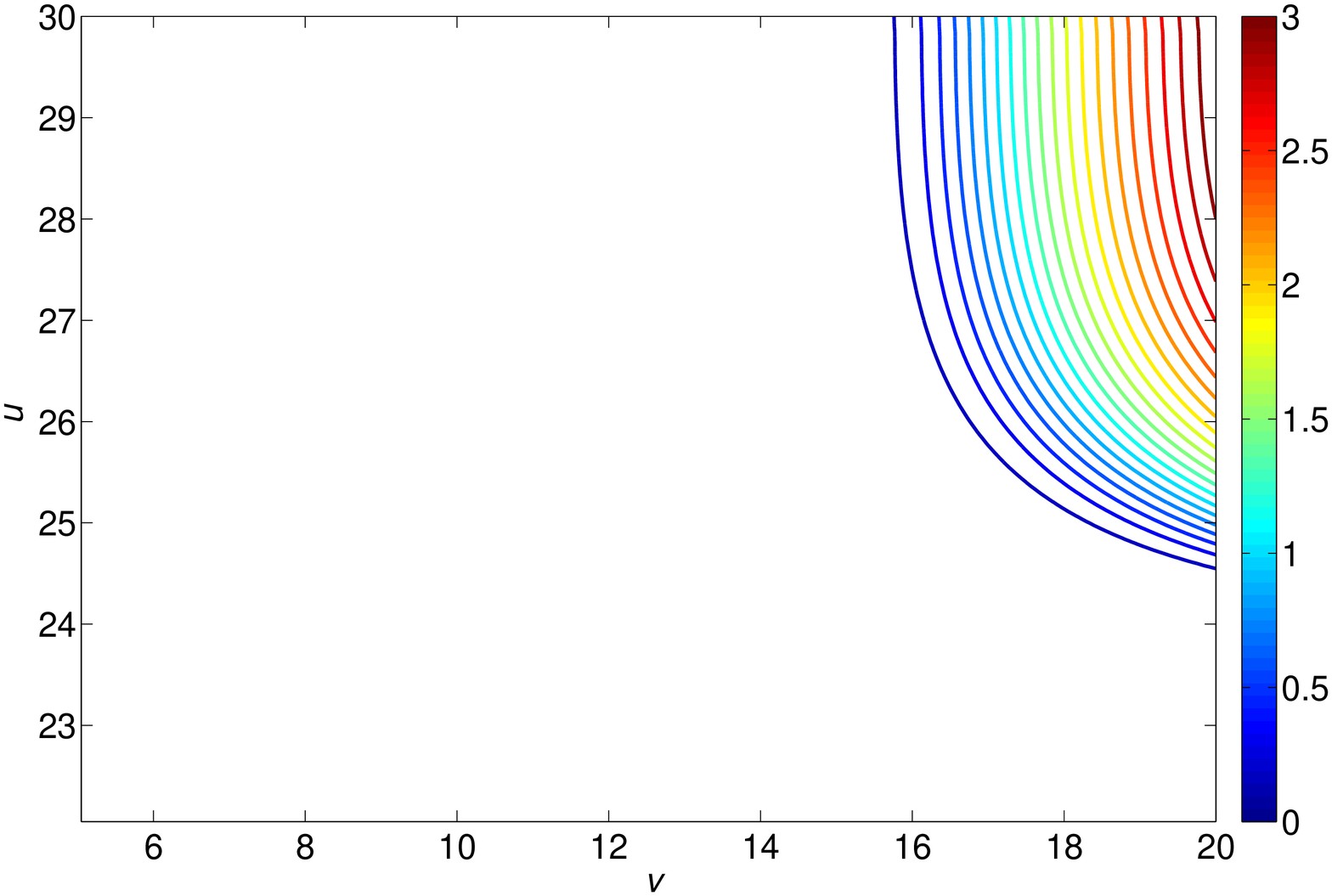}
\includegraphics[width=3.7in,keepaspectratio]{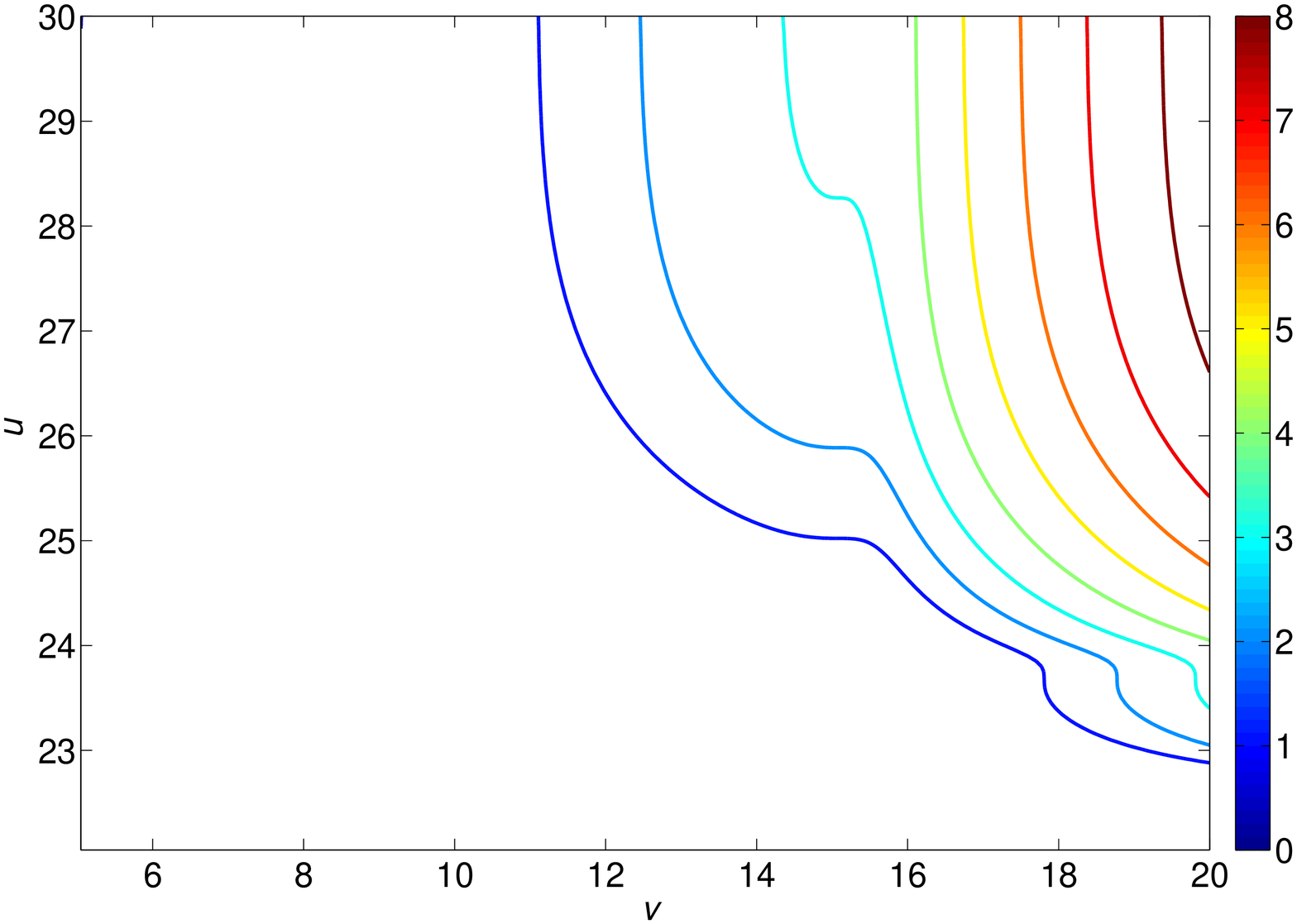}
\caption{\label{figuremass} $\log_{10}(m)$ as a function of $u$ and $v$ for $A=0.05$ (top) and  $A=0.1$ (bottom), showing 
that mass inflation occurs as $v$ increases, approaching the Cauchy horizon of the black hole. The mass function is quite sensitive to $A$.}
\end{figure}

\begin{figure}
\includegraphics[width=3.7in,keepaspectratio]{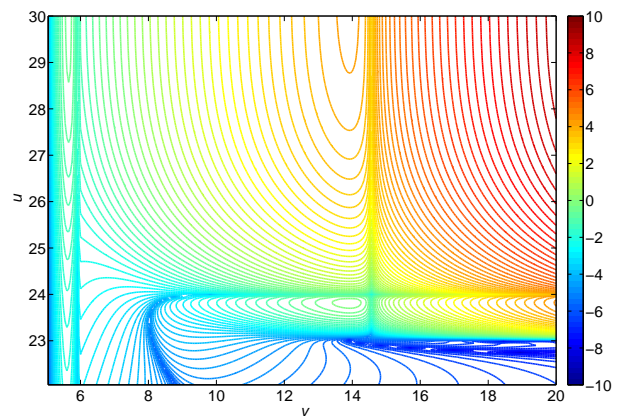}
\caption{\label{curvature} Lines of constant Ricci scalar (actually $\log_{10}|R|$) for $A=0.05$.}
\end{figure}

\begin{figure}
\includegraphics[width=3.6in,keepaspectratio]{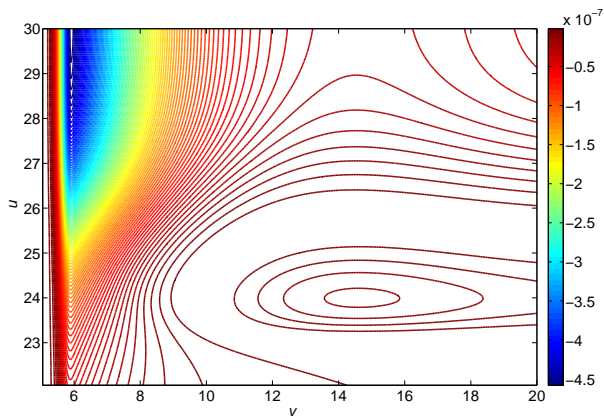}
\caption{\label{figure7c}  $\Phi-1$ as a function of $u$ and $v$ for $A=0.05$. 
The variations of $\Phi$ are small slowly fading away after the initial 
collapse of the scalar field pulse.}
\end{figure}

Fig.~\ref{figuremass} shows $\log_{10}(m)$ as a function of $u$ and $v$, exhibiting the mass 
inflation profile close to the Cauchy horizon of the black hole. These 
figures show an exponential increase of the mass function as $v$ increases.
The increase is consistent with the theorems of \cite{Dafermos:2003wr,Dafermos:2003yw}, which show that, for a compact pulse as here, the mass tends to infinity and a mathematical singularity is attained at the Cauchy horizon, where $v \rightarrow \infty$.
The expectation of a curvature singularity is supported by Fig.~\ref{curvature}, which shows the scalar curvature of the spacetime:
\be
R=\frac{2}{r^2}\left[1+2e^{-2\sigma}\left(\partial_u r\partial_v r+r^2\partial_u\partial_v\sigma+2r\partial_u\partial_vr\right)\right] \ . \ee
The same term
$e^{-2\sigma}\partial_u r\partial_v r$
that appears in (\ref{massinf}) and is responsible for the mass inflation phenomenon
also appears in the Ricci scalar. Thus, not surprisingly, there is a strong growth of the curvature that accompanies the growth of mass, which is manifest in Fig.~\ref{curvature}.

Although one expects that the curvature will diverge only on the Cauchy horizon (at $v\rightarrow+\infty$), which is a null surface, it is important to remark that the curvature should reach Planckian values at finite values of $v$, on a spacelike surface. Beyond this point one expects quantum gravity corrections to become dominant effects, and therefore one cannot trust the geometry given by any classical theory of gravity.

Let us remark that the Cauchy horizon of the eternal RN black hole is composed of two lightlike sheets, as can be seen in Fig.~\ref{figurecp}, one corresponding to $v=+\infty$ and the other to $u=\mbox{constant}$ in our coordinates. The initial conditions imposed create an asymmetry between these two branches of the Cauchy horizon. These conditions are appropriate to see in one branch the mass inflation phenomenon and in the other branch the evolution of the inner horizon under accretion. 
Thus, an outgoing ($u=\mbox{constant}$) observer sees mass inflation (as $v$ increases, cf. Fig. \ref{figuremass}), but an ingoing ($v=\mbox{constant}$) observer sees approximately constant mass, for large $u$.


Finally, Fig.~\ref{figure7c} shows $\Phi-1$ as a function of $u$ and $v$. It shows that the process of mass inflation does not seem to drive rapid variations of the Brans-Dicke scalar. The largest variations of $\Phi$ are associated with the initial collapse of the scalar field pulse onto the black hole. These fluctuations are transient, slowly fading away after the initial collapse of a self-gravitating massless scalar field pulse. The 
amplitude of the fluctuations appears to increase again, slowly, towards the Cauchy horizon. But this growth tends to stop and $\Phi$ to become a function of $r$ only, cf. Fig.~\ref{figure6b} -- top. This is consistent with the homogeneous approximation. 

\subsection{Full non-linear numerical evolution for a non-compact influx}
\label{resultsnoncompact}

\begin{figure}
\includegraphics[width=3.7in,keepaspectratio]{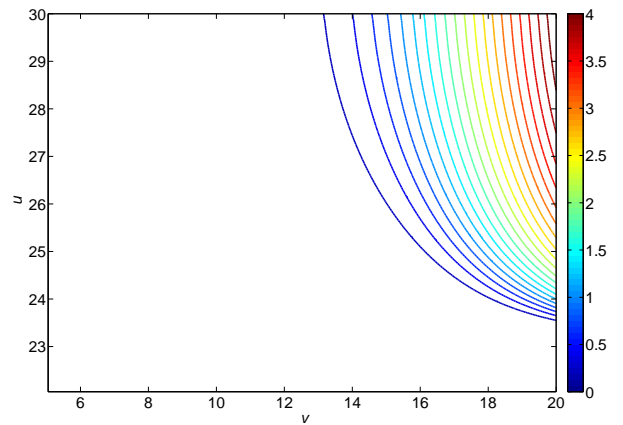}
\caption{\label{noncompact0} Non-compact influx: $\log_{10}(m)$ as a function of $u$ and $v$ for the perturbation \eqref{pulsenc}, with  $A=0.1$. The plot shows that mass inflation occurs close to the Cauchy horizon of the black hole.}
\end{figure}

\begin{figure}
\includegraphics[width=3.7in,keepaspectratio]{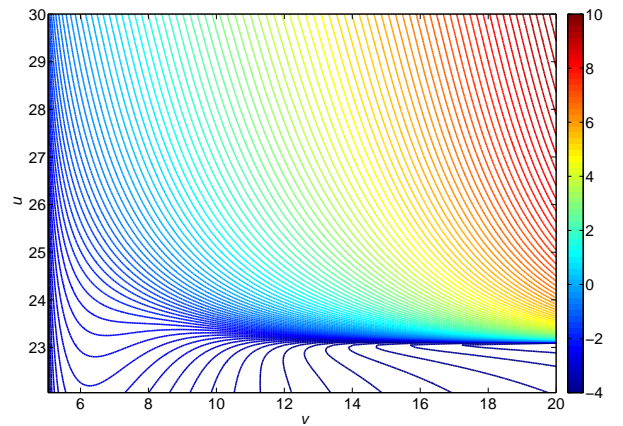}
\caption{\label{noncompact1} Non-compact influx: $\log_{10}|R|$ as a function of $u$ and $v$ for the perturbation \eqref{pulsenc}, with  $A=0.1$. The behaviour of the curvature is qualitatively different from the compact pulse case (cf. Fig.~\ref{figure7c}).}
\end{figure}

\begin{figure}[t!]
\includegraphics[width=3.7in,keepaspectratio]{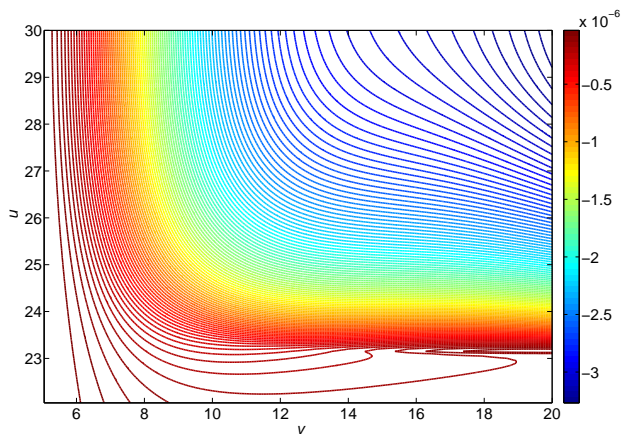}
\caption{\label{noncompact2} Non-compact influx:  $\Phi-1$ for the perturbation \eqref{pulsenc}, with  $A=0.1$. The Brans-Dicke scalar variations are qualitatively different from the compact pulse case (cf. Fig.~\ref{curvature}).}
\end{figure}

The model of the previous subsection,
in which the black hole is fed with a compact pulse of scalar field
over only a finite time, is not realistic.
The model is equivalent to assuming that the black hole remains isolated forever
after accreting the pulse of scalar field.
Real black holes are however never completely isolated,
and they will continue to accrete at some rate or other
until something completely different happens,
like the black hole Hawking evaporating, or the Universe coming to an end
in a big crunch.

In this subsection,
the black hole is fed with an influx of massless scalar field
that starts at some time but then continues into the indefinite future,
equation~(\ref{pulsenc}).

The results of the simulation for the structure of the inner and outer horizon are qualitatively similar to those of the compact pulse. The initial conditions were chosen so as to have a similar black hole mass variation,
of the order of $1\%$ during the time spanned by the simulation,
for both the compact and non-compact influx.
In the latter case, accretion would continue to increase the black hole's mass.

Figs.~\ref{noncompact0} to \ref{noncompact2} show the mass function $\log_{10} m$,
the Ricci scalar $\log_{10} R$, and the variation $\Phi-1$ of the Brans-Dicke scalar.
The behaviour of the mass function, in the integration domain, is similar to the compact pulse case. There is a subtle difference in that ingoing ($v=\mbox{constant}$) observers see more mass inflation than in the compact case. The curvature and Brans-Dicke scalar variations exhibit, on the other hand, a noticeable difference. For the non-compact influx they are monotonic functions of $v$, for large $u$, whereas for the compact pulse, they stop growing at some $v$ and resume growing for larger $v$. The transition seems to be associated to the end of accretion.

\section{\label{conc}Conclusions}

We have studied the internal structure of spherical charged black holes in the framework of the Brans-Dicke theory of gravity, together with the variations of the Brans-Dicke scalar, which determines the gravitational coupling. 

The initial expectations drawn from naive arguments given in the introduction suggested possible large variations of the Brans-Dicke scalar $\Phi$ near the inner horizon of the black hole. The use of the homogeneous approximation indicated that these expectations were correct: in a model with $w=0$, the Brans-Dicke scalar had a divergent behaviour near the inner horizon.  However, in a realistic scenario, accretion will generate counter-streaming and hence mass inflation. In turn, mass inflation changes the dynamics of the Brans-Dicke scalar and prevents large variations of this scalar from occurring. This conclusion was obtained by comparing the behaviour of models with $w=0$ and $w=1$. We emphasise that the non-existence of mass inflation in a single-fluid model with $w=0$ is unrealistic, because even the tiniest admixture of a second fluid would allow relativistic counter-streaming and therefore mass inflation to occur.

One counter-intuitive result drawn from the homogeneous approximation is that mass inflation is more abrupt for a smaller value of the initial perturbation,
in agreement with the conclusions of \cite{Hamilton:2008zz}.

We have performed full non-linear numerical simulations of the evolution of the black hole perturbed by both a compact and a non-compact influx. In the case of the compact pulse, one observes some (expected) changes in the black hole interior, namely the growth of the outer horizon, decrease of one of the light-like sheets of the inner horizon and exponential growth of the effective mass function and curvature near the other light-like sheet of the inner horizon. In this case, the black hole settles down after accretion to a new stationary state, as illustrated in the Carter-Penrose diagram in Fig.~\ref{figurecpaccretion}. The relative variations of the Brans-Dicke scalar are small (between $10^{-6}$ and $10^{-7}$) in the integration region when the compact influx increases the black hole mass by around $1\%$. These variations grow slowly towards the Cauchy horizon, but seem to stabilize. The end of accretion is noticeable, in the case of a compact influx, in both the curvature and Brans-Dicke scalar variations.

For a non-compact influx, on the other hand, the black hole does not settle down to a stationary state. This is a more realistic scenario, since real black holes are not expected to stay isolated forever, after an initial accretion period. For this type of perturbation, the changes in the black hole structure are similar to those of a compact influx. Noticeable qualitative difference appear, however, in the curvature and Brans-Dicke scalar variations. For an accreted mass of around $1\%$ of the black hole mass (in the time spanned by the simulation) these variations are still small, albeit around one order of magnitude larger than for the compact influx.


\begin{acknowledgments}
This work was supported in part by the NSF award AST-0708607. CH is supported by a `Ci\^encia 2007' research contract from Funda\c c\~ ao para a Ci\^encia e a Tecnologia.
\end{acknowledgments}
\bibliography{bd.bib}
\end{document}